\documentclass[11pt]{article}
\usepackage[letterpaper,top=2cm,bottom=2cm,left=3cm,right=3cm,marginparwidth=1.75cm]{geometry}

\usepackage{graphicx}
\usepackage[utf8]{inputenc} 
\usepackage[T1]{fontenc}    
\usepackage[hidelinks, colorlinks=true,citecolor=blue]{hyperref} 
\usepackage{url}
\usepackage{booktabs}
\usepackage{amsfonts}
\usepackage{nicefrac}
\usepackage{microtype}
\usepackage{xcolor}
\usepackage[normalem]{ulem}
\usepackage{amsmath}
\usepackage{amssymb}
\usepackage{color}
\usepackage[numbers]{natbib}
\usepackage{caption}
\usepackage{subcaption}
\usepackage{bm}
\usepackage{bbm}
\usepackage[flushleft]{threeparttable}
\usepackage[shortlabels]{enumitem}
\usepackage{wrapfig}
\setlist{leftmargin=6mm}
\usepackage{multirow}
\usepackage{titlesec}
\usepackage{setspace}

\usepackage{mathtools, mathrsfs}
\usepackage{textcomp, gensymb}
\usepackage{dsfont}
\usepackage{soul}

\usepackage{algorithm}
\usepackage{algorithmic}
\newlength\myindent
\usepackage{amsthm}

\newtheorem{conjecture*}{Conjecture}
\theoremstyle{remark}

\allowdisplaybreaks

\newcommand*{\ie}{\emph{i.e.}{}}

\title{Multi-resolution spatio-temporal prediction with application to wind power generation}

\author{
  Zheng Dong$^1$, Hanyu Zhang$^1$, Shixiang Zhu$^2$\thanks{Email: shixianz@andrew.cmu.edu}, Yao Xie$^1$, and Pascal Van Hentenryck$^1$
}

\date{\small
    $^1$ H. Milton Stewart School of Industrial \& Systems Engineering, Georgia Institute of Technology\\
    $^2$ Heinz College of Information Systems and Public Policy, Carnegie Mellon University
}

\begin{document}

\maketitle

\begin{abstract}
    Wind energy is becoming an increasingly crucial component of a sustainable grid, but its inherent variability and limited predictability present challenges for grid operators. The energy sector needs novel forecasting techniques that can precisely predict the generation of renewable power and offer precise quantification of prediction uncertainty. This will facilitate well-informed decision-making by operators who wish to integrate renewable energy into the power grid.
    This paper presents a novel approach to wind speed prediction with uncertainty quantification using a multi-resolution spatio-temporal Gaussian process. By leveraging information from multiple sources of predictions with varying accuracies and uncertainties, the joint framework provides a more accurate and robust prediction of wind speed while measuring the uncertainty in these predictions. 
    We assess the effectiveness of our proposed framework using real-world wind data obtained from the Midwest region of the United States. Our results demonstrate that the framework enables predictors with varying data resolutions to learn from each other, leading to an enhancement in overall predictive performance. The proposed framework shows a superior performance compared to other state-of-the-art methods.
    The goal of this research is to improve grid operation and management by aiding system operators and policymakers in making better-informed decisions related to energy demand management, energy storage system deployment, and energy supply scheduling. This results in potentially further integration of renewable energy sources into the existing power systems.
\end{abstract}

\noindent%

\section{Introduction}

The role of wind energy in the global energy mix is increasing and is expected to be a crucial element of a more sustainable and greener grid in the coming decades \citep{irena2021wind}. The United States Department of Energy (DOE) predicts that wind energy will surpass other sources of renewable power generation in the next decade \citep{doe2021wind}. However, wind energy sources are known for their inherent variability and limited predictability, which have resulted in significant forecasting errors for some Independent System Operators (ISOs). These forecasting errors present additional challenges for market-clearing algorithms, which are critical for grid operations. As a result, there is a growing need for novel forecasting methods that can accurately {\it quantify} and {\it reduce} the uncertainty of wind power predictions and enable grid operators to make informed decisions \citep{pinson2013wind}.

To manage the increased uncertainty of wind energy sources, one approach is to reduce the prediction resolution by clustering wind farms based on their proximity or estimating average wind conditions over a more extended period. This spatial or temporal aggregation of wind data can not only decrease uncertainty and increase predictability of wind energy but also simplify the training of forecasting algorithms and make downstream market-clearing algorithms more manageable. 
However, oversimplifying the spatio-temporal relationships between wind farms may result in significant losses of wind direction and speed information.
Recent upstream wind conditions, in particular, play a crucial role in predicting future downstream wind power generation. 
Balancing the tradeoff between wind predictions at high and low data resolutions requires more sophisticated models that can account for complex spatio-temporal dependencies between clusters and leverage predictions from different data resolutions.

\begin{figure}[!t]
\centering
\begin{subfigure}[h]{0.19\linewidth}
\includegraphics[width=\linewidth]{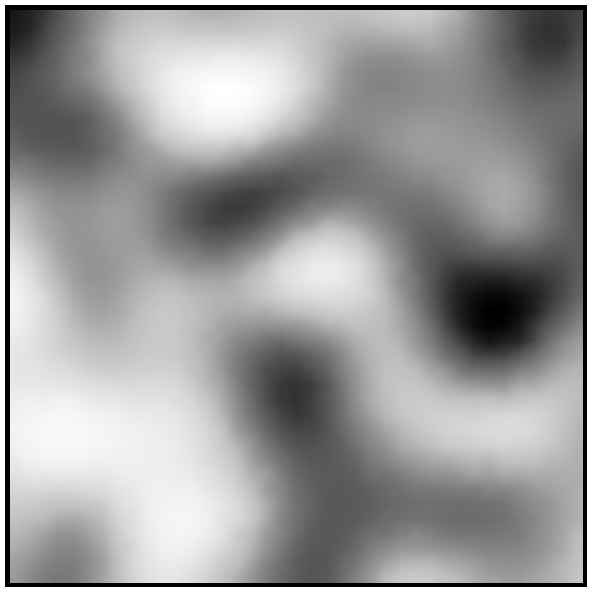}
\caption{$y^*$}
\end{subfigure}
\begin{subfigure}[h]{0.19\linewidth}
\includegraphics[width=\linewidth]{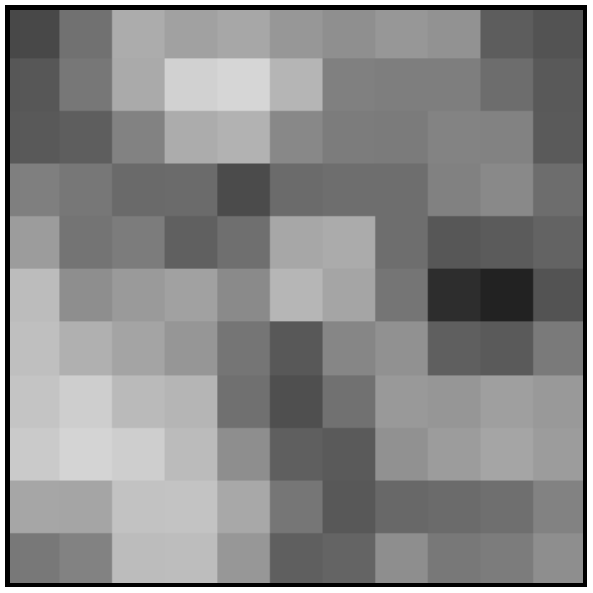}
\caption{$\mathcal{D}_{\xi_1}$}
\end{subfigure}
\begin{subfigure}[h]{0.19\linewidth}
\includegraphics[width=\linewidth]{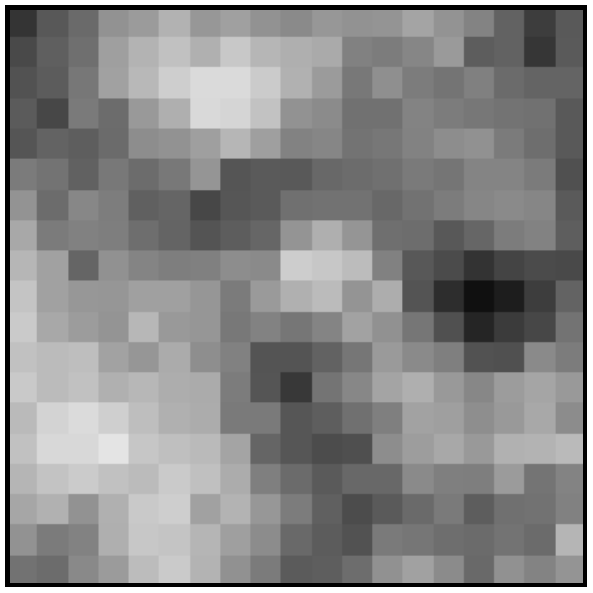}
\caption{$\mathcal{D}_{\xi_2}$}
\end{subfigure}
\begin{subfigure}[h]{0.19\linewidth}
\includegraphics[width=\linewidth]{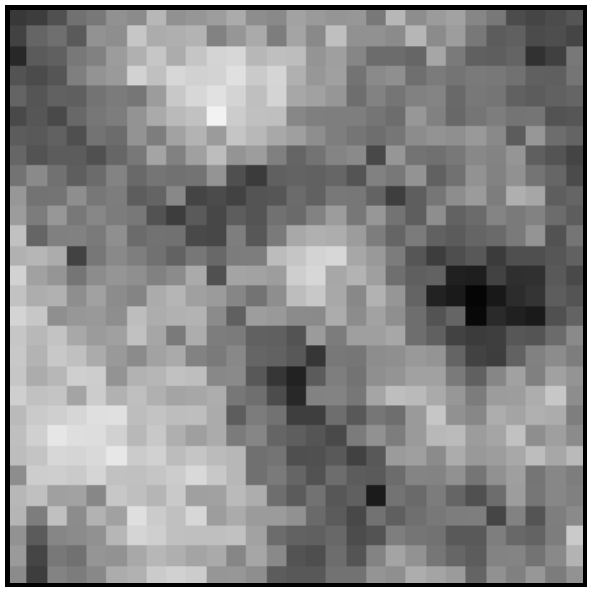}
\caption{$\mathcal{D}_{\xi_3}$}
\end{subfigure}
\begin{subfigure}[h]{0.19\linewidth}
\includegraphics[width=\linewidth]{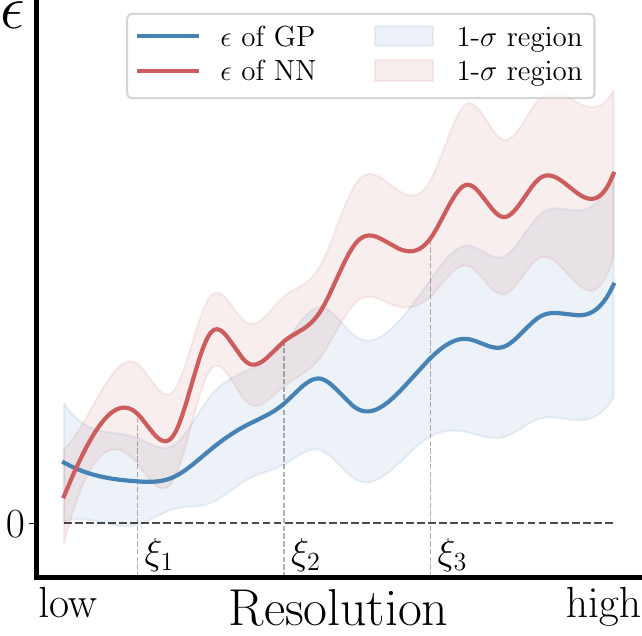}
\caption{$\epsilon$}
\end{subfigure}
\caption{
A motivating example of a two-dimensional Gaussian process at different resolutions.
(a) True process that the data generated by. (b)(c)(d) Aggregation of data marred by a white noise at three resolutions $\xi_1, \xi_2$, and $\xi_3$, respectively.
(d) The prediction error and its uncertainty by a Gaussian process (GP) and a neural network (NN) at different resolutions. Blue and red lines show the evolution of prediction errors of the GP and NN, respectively. The shaded areas indicate the corresponding 1-$\sigma$ region of the prediction error.
}
\label{fig:motivate-example}
\end{figure}

The illustrative example in Figure~\ref{fig:motivate-example} effectively showcases the inherent tradeoff in predictions across varying resolutions.
Models operating at higher resolutions yield predictions rich in data detail, but this comes with increased variability, leading to greater prediction error and uncertainty. On the other hand, models with lower resolutions minimize prediction error and uncertainty by averaging out data fluctuations across broader pixels, albeit at the expense of detail. 
Notably, the error and uncertainty trends in Figure~\ref{fig:motivate-example}(e) change consistently across resolutions, hinting at a possible interrelation between models of varied resolutions. This underlines the importance and potential of crafting frameworks that harness information from multiple resolutions to enhance predictions.

To address the limitations present at specified resolutions and harness the model correlations across multiple resolutions, we develop a multi-resolution model, motivated by multi-fidelity modeling \citep{kennedy2000predicting}, that enables the fusion of knowledge from different sources of predictions with varying accuracy and uncertainties. 
These sources with varying data resolutions are viewed as correlated {\it experts}, with each potentially performing better than the others in specific situations. Additionally, our model assumes that these predictors exhibit smoothness over time and space, meaning that output values for similar spatio-temporal coordinates are reasonably close. 
By capturing the cross-correlation between these sources, we construct a model that can improve predictive performance and quantify uncertainty. 

Specifically, we propose to characterize the correlation of prediction errors across multiple resolutions using a holistic Gaussian process that goes beyond the parametric autoregressive model used for discretized model fidelities \citep{kennedy2000predicting}. 
The Gaussian process is beneficial for estimating values at unobserved locations based on data observed at nearby locations and is flexible to capture complex data dependency.
We adopt the Gaussian process to the joint space of the data domain and resolution space, aiming to provide accurate estimates of the prediction error for wind speed forecasting. The proposed framework has the capability to leverage the knowledge about the covariance structure of the model predictions from multiple resolutions to correct original predictions at each single resolution simultaneously.
Meanwhile, our framework can provide estimates of prediction uncertainty. This is crucial for decision-making processes, as it allows users to assess the reliability of the predictions and make informed choices.
A fine-crafted kernel function is incorporated in the Gaussian process that captures the distinctive covariance patterns of prediction errors at different data resolutions across the domain.
To handle the computational challenge of a large-scale dataset, we leverage sparsity in the model and employ a variational learning strategy for efficient model fitting.

We take advantage of a comprehensive wind data set to demonstrate the effectiveness of the proposed multi-resolution framework in the real-world problem of wind power forecasting.
We present a novel spatio-temporal regressive model with lags that provides accurate wind speed prediction under specified data resolutions. The wind speed is then transformed into power generation through a non-linear mapping \citep{ezzat2019spatio}. We encode the wind direction at each cluster with a directed dynamic graph that captures the physical propagation delay as the wind moves from upstream to downstream. To account for this delay, our model incorporates a carefully crafted correlation function that captures the triggering effects between clusters. The predictive model at each resolution is estimated by minimizing an adaptive mean square prediction error. After the fitted predictive models are obtained, the multi-resolution Gaussian process of prediction error is equipped with a separable kernel that captures the covariance of prediction errors at different spatio-temporal positions and data resolutions and is estimated based on the spatio-temporal prediction results of wind speed. Our numerical results on the wind data demonstrate the good predictive performance of the proposed predictive model and show that the multi-resolution model significantly reduces forecasting errors by correcting the original prediction while providing reasonable uncertainty quantification.

The remainder of this paper is organized as follows. 
After discussing related work, Section~\ref{sec:problem-setup} and \ref{sec:kriging} present the problem setup and our multi-resolution Gaussian process model. The learning strategy for large-scale problems is presented in Section~\ref{sec:variational-learning}.
Section~\ref{sec:wind-data-app} elaborates the technique details of applying the proposed method to a real-world wind data set. In particular, Section~\ref{sec:data} describes and prepares the data sets at multiple resolutions. Section~\ref{sec:regression} presents our novel spatio-temporal regressive model with lags for wind speed prediction. Lastly, Section~\ref{sec:result} presents the numerical results.




\subsection{Related work}

An extensive research effort has been devoted to wind prediction (e.g., \cite{giebel2011state, LEI2009915, soman2010review}). 
Early attempts \citep{lange2008new, potter2006very} resort to physical models, relying on parameterizations based on a detailed physical description of the atmosphere. For instance, the Numeric Weather Prediction (NWP) usually runs 1 or 2 times per day due to the difficulty and cost of acquiring real-time information, which limits its usefulness to medium to long-term forecasts ($> 6$ hours ahead) \citep{LEI2009915, soman2010review}. \cite{Kosovic2020} mentions that ``the best strategies for nowcasts (0- to about 3-hour ahead) rely on observations near the wind farm.''
In addition, the parameterization of the physical models requires critical information about the surrounding atmosphere, which could inject bias into the prediction \citep{tascikaraoglu2014review}.
To provide timely and reasonable predictions, many statistical methods are proposed \citep{brown1984time, erdem2011arma, he2014spatio, kusiak2010estimation, pourhabib2016short} to predict the wind speed or power using past observations.
For instance, time series models \citep{chen2013garch, torres2005forecast, yatiyana2017wind, zhou2010numerical} have been widely used to provide short-term wind forecasting by aggregating historical wind data at fixed time scale as time sequences.
Another line of work for wind prediction \citep{babazadeh2012hour, chen2014short} focuses on Kalman filter, improving forecast precision at a minute- or hour-level resolution through dynamically modified prediction weights. 
Recent work has focused on machine-learning models for wind prediction \citep{CHEN2020491, mohandes2004support, SALINAS20201181} to take advantage of their capability for modeling nonlinear dependencies, such as support vector machine (SVM) \citep{mohandes2004support, mi2019wind, yang2015support} and recurrent neural networks \citep{yao2018multidimensional, yu2019lstm}.
The previously mentioned techniques mostly offer wind forecasting at a coarse, aggregated data resolution without fully harnessing the hierarchical insights embedded within the raw data. A few of these methods \citep{chen2013garch, yao2018multidimensional, yu2019lstm} strive to predict wind speed and power at the native high data resolution but fail to manage the uncertainty associated with their predictions.

Various techniques for multi-resolution wind power forecasting have emerged \citep{chen2020medium, chen2023multi, doucoure2016time, he2020long, liu2020corrected, liu2020wind, nejati2023new} to exploit the hierarchical relationships present within wind data across different levels of detail. In the study by \cite{chen2023multi}, neural networks have been adopted to extract data features from data aggregated at different levels. This process enables the subsequent creation of wind predictions at a specific level using the acquired features. Another research avenue centers on the utilization of ensemble models, as demonstrated by \cite{chen2020medium} and \cite{liu2020corrected}, to yield improved final predictions surpassing the outcomes of individual models operating at singular resolutions. These ensemble models involve multiple stages and components designed to achieve diverse objectives in wind prediction, error correction, and uncertainty quantification. On the contrary, this study focuses on the establishment of a unified model that is capable of capturing the multi-resolution correlation between different levels of predictions and providing corrections for predictions at each single resolution. A noteworthy distinction lies in the fact that wind predictions in our framework operates without being tied to a specific model. The concept of employing artificial wavelet neural networks in time series predictions \citep{doucoure2016time} exhibits a similar modeling capacity for enhancing wind predictions across multiple resolutions as in this study. Nevertheless, their model primarily targets the reduction of complexity in multi-resolution prediction while maintaining the predictive accuracy, unlike the primary objective of multi-resolution forecasting correction in our study.

Our proposed method introduces a holistic Gaussian process to provide error correction and manage uncertainty by considering their correlations over time, space, and data resolutions. The idea is motivated by multi-fidelity models introduced by \cite{kennedy2000predicting}.
Multi-fidelity models have been commonly used in many scientific and engineering problems \citep{de2015uncertainty, forrester2007multi, kennedy2000predicting, liu2022multi}. They provide tools for model calibration and confidence interval construction by exploiting the correlation between different fidelity levels. For example, they are adopted in \cite{jones1998efficient, venter1998construction} to enhance the quality of the cheap approximations from low-fidelity models to unknown response surfaces with the information of expensive high-fidelity models.
In general, multi-fidelity models involve conducting experiments at varying levels of complexity. In this particular study, we consider the aggregation and forecasting of wind data at different resolutions as analogous processes.
We adopt a Bayesian approach, where our initial beliefs regarding the correlations of prediction errors across various resolutions are captured using a Gaussian process. This framework enables us to assess the amount of bias to be corrected and effectively manage predictive uncertainty in unobserved situations. The major difference lies in that while multi-fidelity models primarily concentrate on refining the accuracy of lower-fidelity models through insights from higher-fidelity models, our objective centers on elevating the predictive accuracy and quantifying the predictive uncertainty of models at individual resolutions. This is achieved by simultaneously leveraging error information gathered from multiple resolutions.

Finally, a few studies investigated uncertainty quantification of wind power forecasts \citep{quan2019survey, yan2015reviews}. 
Traditional risk index methods intuitively provide information for uncertainty quantification from a physical perspective \citep{pinson2004line, pinson2007methods}. They quantify the uncertainty in prediction by considering a collection of influential factors about weather and atmospheric system and their relationship with the prediction error. However, due to the instability of the atmospheric system, the uncertainty of the prediction is high, making it prone to errors. 
There are other attempts in scenario simulation \citep{aghaei2013scenario, li2016scenario, pinson2007generation} to quantify uncertainty by generating scenarios of wind power generation that are close enough to the real observation. 
However, it would be computationally intensive to generate a large number of scenarios in a bid to attain accurate uncertainty quantification. 
Some probabilistic methods have been developed to quantify uncertainty for wind power prediction. By assuming a parametric form for the wind, a lower and upper bound of prediction intervals can be easily obtained \citep{khosravi2014optimized, pinson2006estimation}. 
Despite being computationally efficient, the probabilistic models could be misspecified and lack generalization to other data.

\section{Proposed framework}
\label{sec:proposed-framework}

\subsection{Problem setup}
\label{sec:problem-setup}


Consider the collection of wind speed observations within a continuous spatio-temporal domain $\Omega \coloneqq \mathscr{T} \times \mathscr{S} \subseteq \mathbb{R}^3$. The generation of wind speed can be described by the following model:
\[
    \widetilde{y}(t, s) = y^*(t, s) + w(t, s),
\]
where $y^*(t, s)$ is a random field over space, and $w(t, s)$ is white noise.
In practice, the wind speed is observed under a designated data \emph{resolution}, where the collected data point at time $t$ and location $s$ is the result of grouping or averaging the wind speed variable over a specified spatio-temporal neighborhood that contains the point $(t, s)$.
To provide tangible examples, the resolution could symbolize the sampling frequency for a time series or the count of spatial subregions. We involve a complexity parameter $\xi$ to represent each data resolution and assume that we have observations across a finite set of data resolutions $\{\xi_1, \xi_2, \dots\} \subset \Xi$, where $\Xi$ is the resolution space.
Under each resolution $\xi_i$, the entire spatio-temporal domain is discretized into a finite set of subregions, denoted as $\{\Omega_{i, j}\}_{j \in \mathcal{J}_i}$ where $\mathcal{J}_i$ represents the set of subregion indices. The $\{\Omega_{i, j}\}_{j \in \mathcal{J}_i}$ form a partition of $\Omega$, \ie, $\cup_{j \in \mathcal{J}_i}\Omega_{i, j} = \Omega$ and $\Omega_{i, j} \cap \Omega_{i, j'} = \varnothing, \forall j \neq j'$. Each subregion $\Omega_{i, j}$ is represented by its centroid $x_{i, j} = (t_{i, j}, s_{i, j})$. The corresponding observation associated with each subregion is the outcome of an unknown aggregation function $Q$, denoted as $y_{i, j} = Q(\widetilde{y}, \Omega_{i, j})$. As an example, one possible form of the aggregation function is $Q(\widetilde{y}, \Omega_{i, j}) = \iint_{\Omega_{i, j}}\widetilde{y}(t, s)dsdt / |\Omega_{i, j}|$.

The traditional task of wind power prediction involves constructing a predictive model $f_i$ that takes different times and locations as input and generates wind speed predictions as output at the specified data resolution $\xi_i$. 
The observations acquired at resolution $\xi_i$ are collectively denoted as $\mathcal{D}_i \coloneqq \{(x_{i, j}, y_{i, j})\}_{j\in \mathcal{J}_i}$.
The fitted predictive model, denoted as $\widehat f_i$, is obtained using the dataset $\mathcal{D}_i$ by minimizing a predefined loss $L$. More formally, 
\[\widehat f_i \coloneqq \underset{f_i \in \mathcal{F}}{\arg \min}\ L(f_i; \mathcal{D}_i).\]

In various engineering problems \citep{hanna2017coarse, phillips1973strategy, resseguier2021new, skamarock1989adaptive}, it is common to make predictions in a single and coarser domain. This practice can enhance prediction accuracy and effectively handle uncertainty by carefully choosing an appropriate data resolution.
However, low-resolution models may lack the precision required to capture true dynamics, especially when certain crucial data patterns are oversimplified. 
To address this challenge and draw inspiration from the concept of multi-fidelity models introduced by \cite{kennedy2000predicting}, we propose a novel solution in the form of a surrogate model. 
We aim to capture predictive discrepancies by considering the correlation of predictions across multiple resolutions.
Our approach relies on the key assumption that strong correlations exist in predictive errors when two data resolutions are similar. 
Rather than relying on a single predictive model at a specific data resolution, we leverage multiple predictive models across various data resolutions and tap into richer knowledge about the underlying data dynamics. 

Formally, the relationship between the observation and the corresponding prediction at resolution $\xi_i$ can be expressed as
\begin{equation}
    y = \widehat f_{i}(x) + \epsilon_i,~\quad~\text{where}~x \in \Omega~\text{and}~y \in \mathbb{R}.
    \label{eq:proposed-framework-of-variable}
 \end{equation}
Here the predictive error $\epsilon_i$ is introduced to capture the discrepancy between the observed response $y$ and the prediction $\widehat f_{i}$. 
Our goal is to jointly model $\epsilon$ from multiple resolutions, which can be used to correct predictions for all resolutions and enable accurate uncertainty quantification for these predictions.




\subsection{Multi-resolution Gaussian process}
\label{sec:kriging}

This section presents our corrected prediction based on a holistic Gaussian process (GP) that aims to capture the prior belief about the relationship of each level of the predictions.
In practical applications, it becomes important to account not only for the correlation of prediction errors across resolutions within $\Xi$ but also the correlations of errors across different subregions, based on their respective distances.
To this end, we represent the $j$-th subregion at resolution $\xi_i$ using a tuple $u = (i, j) \in \mathscr{U}$ that can be regarded as a spatio-temporal-resolution \emph{coordinate} in the studied space $\mathscr{U} \subseteq \mathcal{I} \times \mathcal{J}$,
where $\mathcal{I}$ is the set of resolution indices and $\mathcal{J} \coloneqq \cup_{i \in \mathcal{I}}\mathcal{J}_i$ is the aggregated collection of subregion indices at resolutions within $\mathcal{I}$.
In light of this, we extend the definition of the error term $\epsilon$ to take into consideration both the resolution and the data domain, resulting in a error function $\epsilon(u)$ in terms of the coordinate.
We then use a Gaussian process (GP) with a zero mean to characterize such complex dependencies of prediction errors, where the covariance structure of this GP is specified by a kernel function represented as $k$. 
Therefore, the joint distribution of the errors collected at any set of $N$ coordinates, denoted as $U \subseteq \mathscr{U}$, can be characterized by
\begin{equation}
    p(\epsilon) = \mathcal{N}(\epsilon~|~
    0, K_{UU}).
    \label{eq:noise}
\end{equation}
Here $\epsilon \coloneqq \{\epsilon(u)\}_{u \in U}$, $K_{UU}$
is an $N \times N$ matrix and its entries are pairwise evaluations of the kernel function $k(u, u')$, $\forall u, u' \in U$.
The $\mathcal{N}(\cdot|\mu, \Sigma)$ is the probability density function of a multivariate Gaussian distribution with mean $\mu$ and variance $\Sigma$.
The parameters of GP can be optimized by maximizing the log marginal likelihood of the observed data. We further denote the collection of observations and model predictions at coordinate set $U$ as $y \coloneqq \{y_u\}_{u \in U}$ and $\widehat f \coloneqq \{\widehat f_i(x_u)\}_{u \in U}$, respectively.
Under the assumption that the predictions $\widehat f$ are deterministic where predictive models have been previously estimated from observed data in a separate stage, the conditional probability of the observed data $y$ can then be expressed as $p(y) = \mathcal{N}(y~|~\widehat f, K_{UU})$.
The model parameters are estimated by solving:
\begin{equation}
\begin{aligned}
    \underset{\theta \in \Theta}{\arg \max}~\ell(\theta) 
    \coloneqq \log p(y) = -\frac{1}{2} ( y - \widehat f )^\top K_{UU}^{-1} ( y - \widehat f ) - \frac{1}{2} \log |K_{UU}| - \frac{N}{2} \log 2\pi,
    \label{eq:objective}
\end{aligned}
\end{equation}
where $\theta$ denotes the set of model parameters and $\Theta$ is the corresponding parameter space.

\paragraph{Kernel design} It is worth noting that the choice of the kernel function $k$ in our framework plays a key role in describing the data dynamics. 
Previous work of Simple Kriging \citep{olea2012geostatistics} uses a stationary kernel function, assuming that the covariance between data points relies on their distance. 
However, this assumption might prove inadequate in scenarios where heterogeneous correlations dominate the input space, especially in cases of longitudinal and spatial data.
In addition, incorporating data correlations across various resolutions demands deeper exploration in crafting suitable kernel designs.
This is exemplified by the observation that data at higher resolutions often exhibit significantly larger variance compared to those at lower resolutions. 
This underscores the necessity for a meticulous choice of kernel functions that can adapt to the unique behaviors inherent in the data.
Last but not least, maintaining an invertible Gram matrix $K_{UU}$ holds critical importance in upholding the integrity of the framework.

In practice, we choose a kernel design that assumes the time, space, and data resolution are mutually independent, as each factor describes the wind power characteristics within distinct domains. The kernel function at coordinates $u=(i, j)$ and $u'=(i', j')$ is thus separable, i.e.,
\begin{align*}
    &~k\left(u, u' \right) = \upsilon_s(s_{i, j}, s_{i', j'}) \cdot \upsilon_t(t_{i, j}, t_{i', j'}) \cdot \phi(\xi_i, \xi_{i'}).
\end{align*}
The kernels of time and space coordinates are assumed to be commonly used Gaussian correlation functions:
$
    \upsilon_s(s_{i, j}, s_{i', j'}) = \exp\left\{ - \sigma_s ||s_{i, j} - s_{i', j'}||^2 \right\},
    \upsilon_t(t_{i, j}, t_{i', j'}) =\exp\left\{ - \sigma_t (t_{i, j} - t_{i', j'})^2 \right\},
$
where $\sigma_s > 0$ and $\sigma_t > 0$ are two learnable parameters. 
Design of $\phi(\xi_i, \xi_{i'})$ requires more deliberation.
The function value would become larger spatio-temporally when the resolution increases as shown in Fig.~\ref{fig:pred-out-of-sample}, the following correlation function for both $\phi(\xi_i, \xi_{i'})$ can be selected:
$
    \phi(\xi_i, \xi_{i'}) =
    \exp\left\{ - \sigma^\xi_0 \|\xi_i - \xi_{i'}\|^2 \right\} + \exp\left\{-\sigma^\xi_1/\|\xi_i\|^2 \right\} +
    \exp\left\{-\sigma^\xi_1/\|\xi_{i'}\|^2 \right\},
$
where $\sigma^\xi_0, \sigma^\xi_1 > 0$ are learnable parameters. Generally, the resulting Gram matrix $K_{UU}$ is invertible because the time, space and data resolution are not linearly dependent.

\subsection{Variational learning}
\label{sec:variational-learning}
The GP approach is notoriously intractable for large datasets since the computations require the inversion of a matrix of size $N \times N$ which scales as $O(N^3)$ \citep{rasmussen2003gaussian}. 
This challenge becomes even more pronounced when dealing with multi-resolution predictions, as a multitude of observations across various data resolutions further compounds the issue.
To address the tractability issues, this paper derives sparse models for the error $\epsilon$ inspired by previous research \citep{hensman2013gaussian, hensman2015scalable, titsias2009variational, zhu2021early}. 
The idea is to introduce a small set of $M$ auxiliary inducing variables $\{z(v)\}_{v \in V}$ evaluated at a set of inducing points $V \subset \mathscr{U}$ that aims to best approximate the training data. The initial inputs $V$ are a subset of coordinates randomly sampled from $\mathscr{U}$. It is then possible to adopt a variational learning strategy for such a sparse approximation and jointly infer the optimal inducing inputs and other model parameters by maximizing a lower bound of the true log marginal likelihood of the observed data $y$.

Inducing variables $z$ are drawn from the same GP prior of $\epsilon$ in \eqref{eq:noise}, hence the joint distribution can be written as
\begin{equation}
    p(\epsilon, z) = \mathcal{N}\left(~
    \begin{bmatrix}
    \epsilon \\
    z
    \end{bmatrix}
    ~\bigg|~
    0,~
    \begin{bmatrix}
    K_{UU} & K_{UV} \\
    K_{UV}^\top & K_{VV}
    \end{bmatrix}~
    \right),
    \label{eq:joint-dist-f-u}
\end{equation}
where $K_{VV}$ is formed by evaluating the kernel function pairwisely at all pairs of inducing points in $V$, and $K_{UV}$ is formed by evaluating the kernel function across the data points $U$ and inducing points $z$. 

To obtain a computationally efficient inference, the posterior distribution $p(\epsilon, z|y)$ over random variable vector $\epsilon$ and $z$ is approximated by a variational distribution $q(\epsilon, z) \coloneqq p(\epsilon | z) q(z)$. 
To jointly determine the variational parameters and model parameters,
the variational evidence lower bound (ELBO) \citep{hoffman2016elbo}
substitutes for the marginal likelihood defined in \eqref{eq:objective}:
\begin{equation}
    \log p(y) \ge \mathbb{E}_{q(\epsilon)}\left [ \log p(y|\epsilon) \right ] - \text{KL}\left [ q(z) || p(z) \right ],
    \label{eq:elbo}
\end{equation}
where $\text{KL}[q||p]$ denotes the Kullback–Leibler (KL) divergence between two distributions $q$ and $p$ \citep{kullback1951information}. The derivation defines $q(\epsilon) \coloneqq \int p(\epsilon|z) q(z) dz$ and assumes $q(z) \coloneqq \mathcal{N}(z | m, S)$, which is the most common way to parameterize the prior distribution of inducing variables in terms of a mean vector $m$ and a covariance matrix $S$.
To ensure that the covariance matrix remains positive definite, it can be represented as a lower triangular form $S = L L^\top$. This leads to the analytical form for $q(\epsilon)$:
\[
    q(\epsilon) = \mathcal{N}(\epsilon~|~Am, K_{UU} + A (S - K_{VV}) A^\top),
\]
where $A = K_{UV} K_{VV}^{-1}$. 
The likelihood can also be factorized as $p(y|\epsilon) = \prod_{n=1}^N p(y_n|\epsilon_n)$ for the ease of computation in \eqref{eq:elbo}. Therefore, the ELBO objective can be rewritten as
\begin{equation}
\begin{aligned}
    &~\ell_\text{ELBO}(V, m, S) \coloneqq \sum_{n=1}^N \mathbb{E}_{q(\epsilon_n)}\left [ \log p(y_n|\epsilon_n) \right ] - \text{KL}\left [ q(z) || p(z) \right ].
    \label{eq:obj-elbo}
\end{aligned}
\end{equation}
Note that the one-dimensional integrals of the log-likelihood in \eqref{eq:obj-elbo} can be computed by Gauss-Hermite quadrature \citep{liu1994note} (the derivation of the ELBO can be found in Appendix~\ref{append:elbo}).
In contrast to directly maximizing the marginal log-likelihood defined in \eqref{eq:objective}, computing this objective and its derivatives only costs $O(NM^2)$. 
In practice, the optimization is carried out through stochastic gradient descent (see Appendix~\ref{append:sgd}).

\subsection{Model correction and uncertainty quantification} 
Our proposed framework is capable to improve predictive accuracy and manage uncertainty at every individual resolution.
Under a specific data resolution $\xi_i \in \Xi$, we assume that the observations $\{y_u\}_{u \in U^o}$ and predictions $\{\widehat f_i(x_u)\}_{u \in U^o}$ have been obtained at a subset of coordinates $U^o \coloneqq \{(i, j)\}_{j \in \mathcal{J}_{i}^{o}}$, where $\mathcal{J}_{i}^{o}$ represents the index set of those observed subregions. We are interested in correcting predictions at unobserved coordinates $U^* \coloneqq \{(i, j)\}_{j \in \mathcal{J}_{i}-\mathcal{J}_{i}^{o}}$.
This is achieved by leveraging the variational posterior of prediction error at $U^*$ estimated by the fitted GP and inducing point $V$. 
The posterior distribution of the error $\epsilon^*$ at $U^*$ is given by
\begin{equation}
    p(\epsilon^*|\{y_u\}_{u \in U^o}) 
    = \mathcal{N}(\epsilon^*~|~A_* m, A_* S {A_*}^\top + B_*),
    \label{eq:pred-posterior}
\end{equation}
where $A_* = K_{*V}K_{VV}^{-1}$ and $B_* = K_{**} - K_{*V} K_{VV}^{-1} K_{*V}^\top$.
The $K_{* V}$ denotes a $|U^*| \times M$ matrix and its entries are pairwise evaluations of $k(u^*, v)$ where $u^* \in U^*$ and $v \in V$. 
The derivation of the predictive posterior can be found in Appendix~\ref{append:pred-posterior}.
The posterior mean of $\epsilon^*$ is an estimation of prediction errors at unobserved coordinates $U^*$, thus the corrected predictions can be obtained using \eqref{eq:proposed-framework-of-variable}, \ie, by adding the posterior mean to the original predictions $\widehat f^* \coloneqq \{\widehat f_i(x_u)\}_{u \in U^*}$. The uncertainty of the corrected predictions can be quantified by the posterior variance of $\epsilon^*$.

\section{Case study: Wind speed prediction}
\label{sec:wind-data-app}

In this section, we demonstrate the capability of the proposed framework in the practical engineering problem of multi-resolution spatio-temporal wind speed prediction, and assess the effectiveness of the framework in enhancing the predictive performance and managing the model uncertainty using real-world wind data from the Midwest region of the United States. After introducing and preparing the multi-resolution wind data, we present a novel spatio-temporal regressive architecture for the predictive model $f_\xi$, which is fitted and provides accurate wind speed predictions under each specified resolution. 
Comprehensive experimental results showing the superiority of our proposed framework are documented.

\subsection{Data overview}
\label{sec:data}


\begin{figure}[!t]
\centering
\begin{subfigure}[h]{0.35\linewidth}
\includegraphics[width=\linewidth]{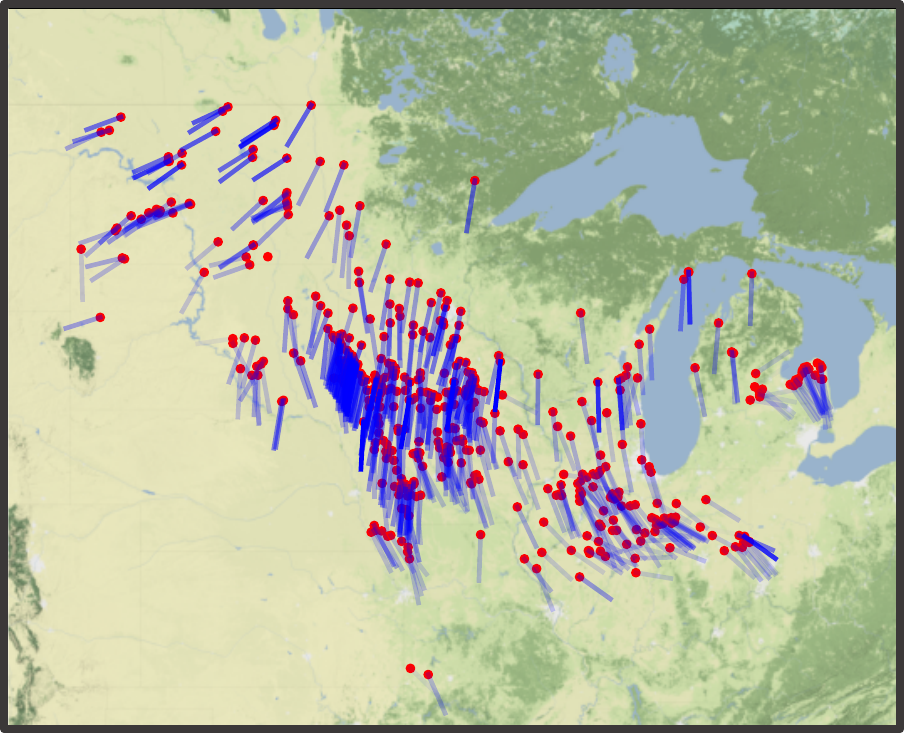}
\caption{12:00 AM, Sept 21, 2020}
\end{subfigure}
\hspace{0.1in}
\begin{subfigure}[h]{0.35\linewidth}
\includegraphics[width=\linewidth]{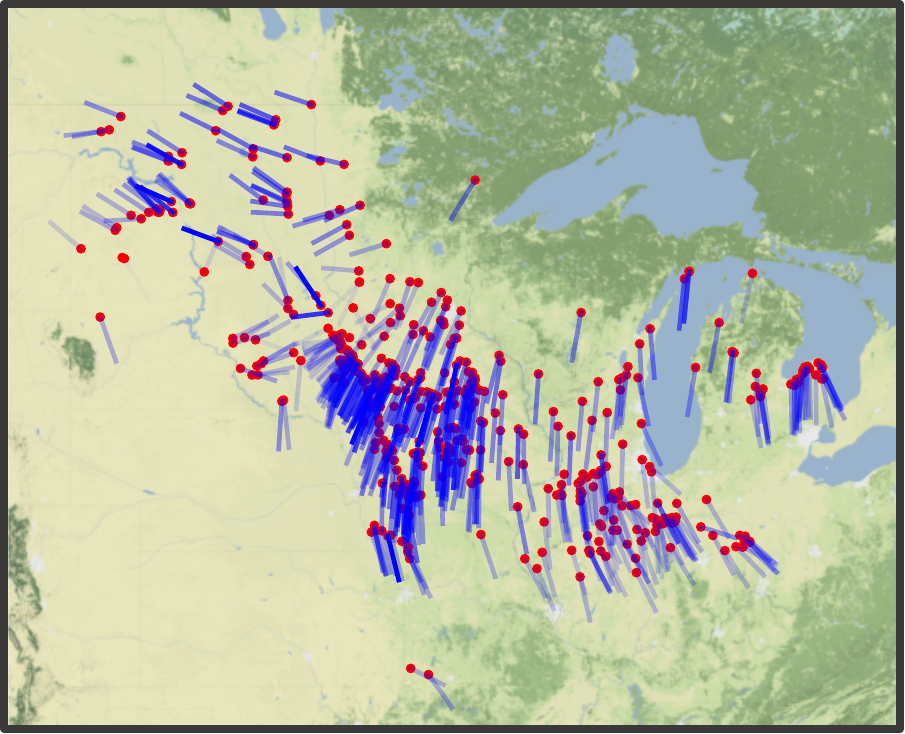}
\caption{12:00 PM, Sept 21, 2020}
\end{subfigure}
\caption{Two snapshots of the real wind data: 506 wind farms in the Midwestern United States. The dots indicate the location of these farms. The lines indicate the wind directions, and the color depth represents the wind speeds. Geographic coordinates of the wind farms have been concealed to maintain confidentiality.} 
\label{fig:raw-data}
\end{figure}

\begin{figure}[!t]
\centering
\begin{subfigure}[h]{.24\linewidth}
    \includegraphics[width=\linewidth]{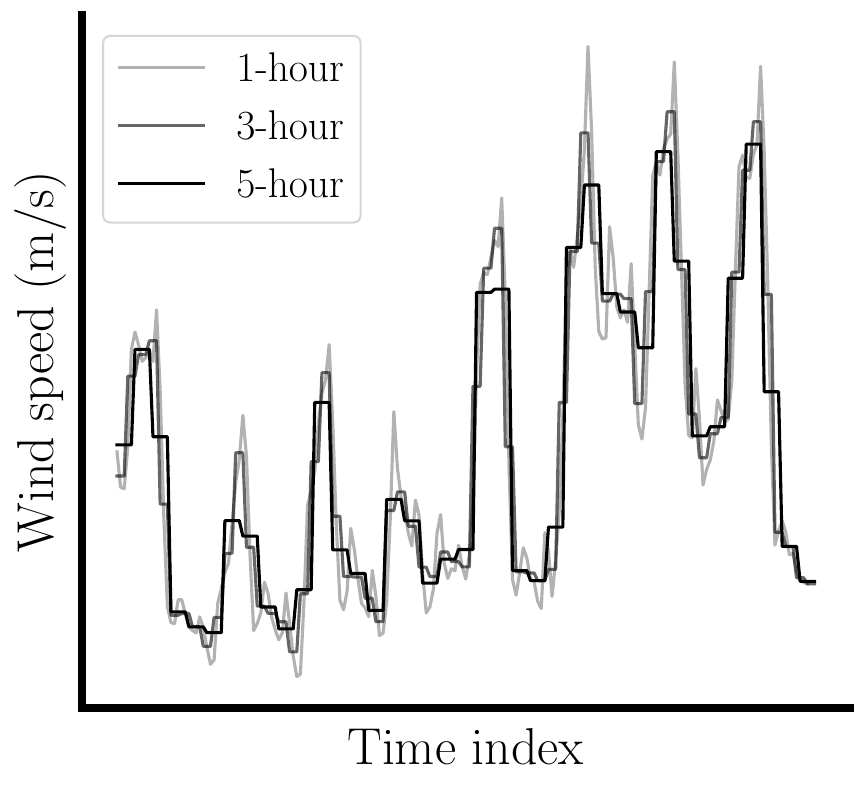}
\caption{Wind speed}
\end{subfigure}
\begin{subfigure}[h]{.24\linewidth}
    \includegraphics[width=\linewidth]{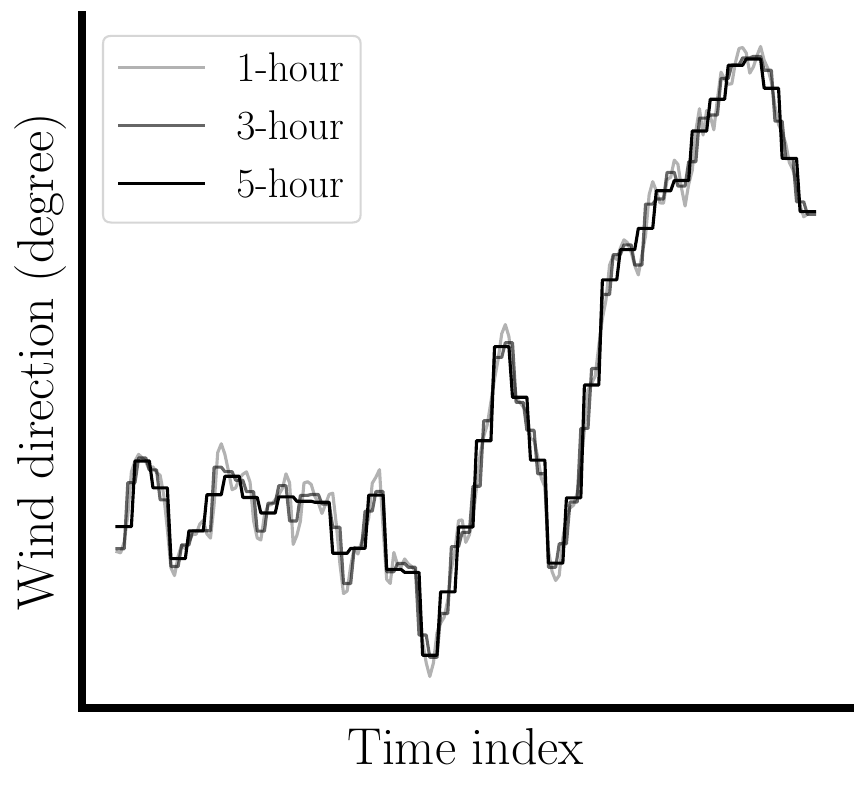}
\caption{Wind direction}
\end{subfigure}
\begin{subfigure}[h]{0.24\linewidth}
\includegraphics[width=\linewidth]{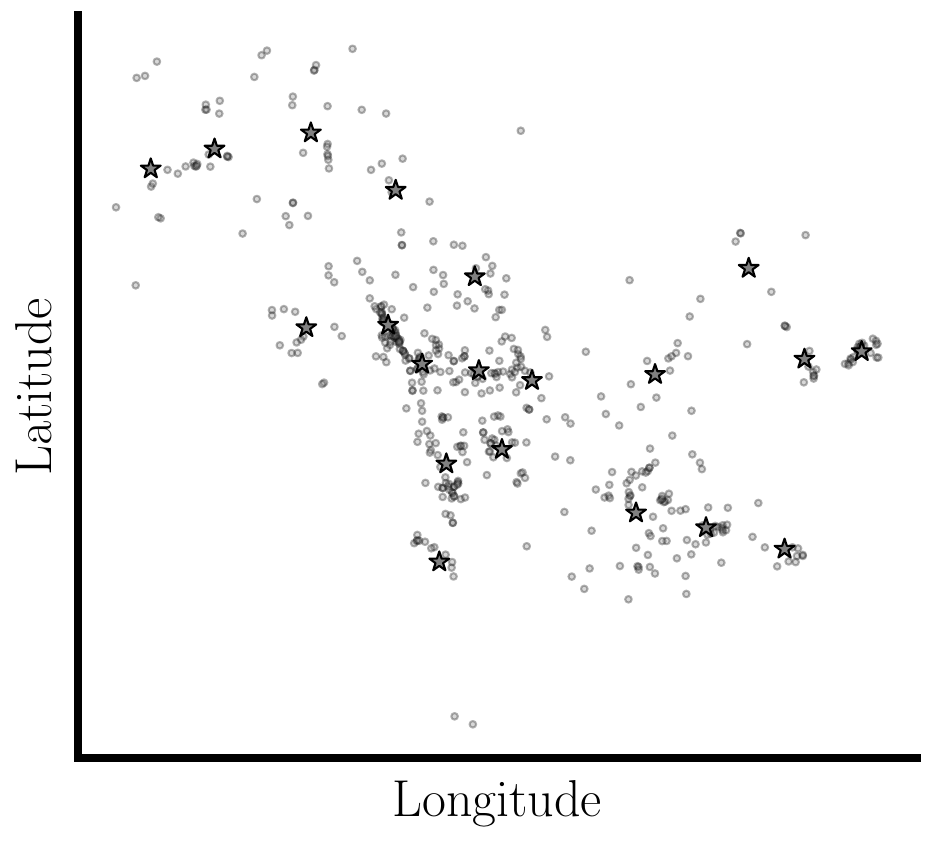}
\caption{20 clusters}
\end{subfigure}
\begin{subfigure}[h]{0.24\linewidth}
\includegraphics[width=\linewidth]{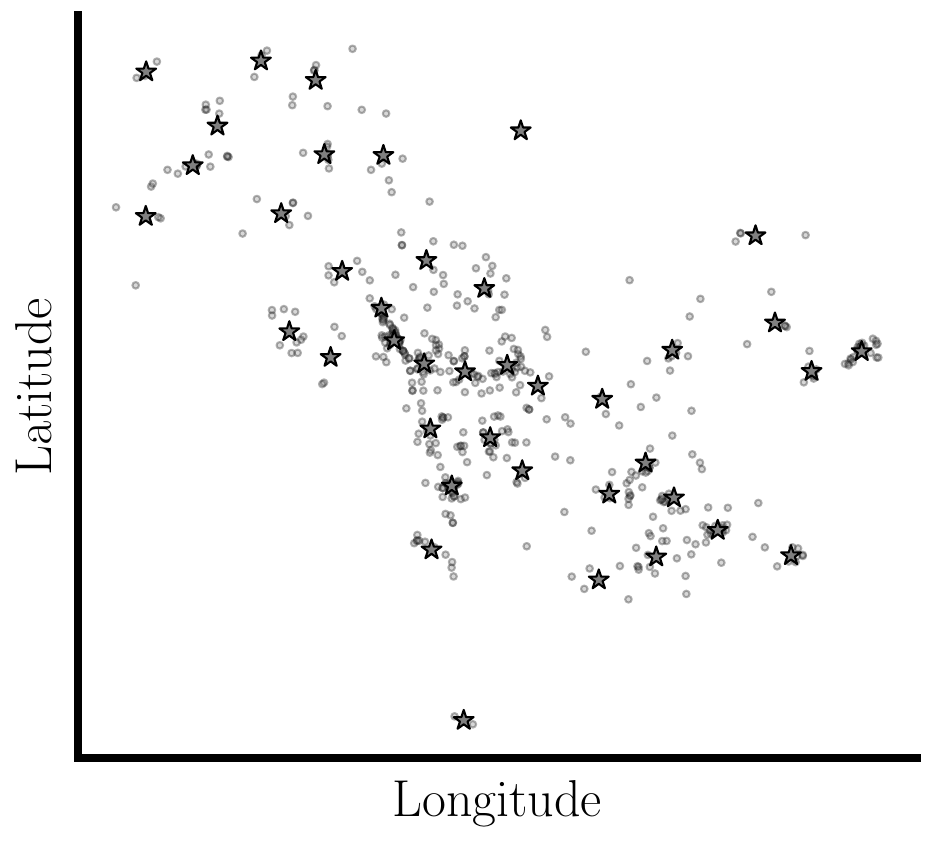}
\caption{40 clusters}
\end{subfigure}
\caption{
(a)(b) Wind speed and wind direction at three time resolutions: $\xi_t=T/4$ (1-hour), $\xi_t = T/12$ (3-hour), and $\xi_t=T/20$ (5-hour). (c)(d) Wind farm groups at two space resolutions. Farms (dots) are grouped according to their proximity using the $k$-means algorithm, with each group centroid represented by a star marker.
}
\label{fig:map-diff-spatial-resolution}
\end{figure}

The wind data in this study include 506 wind farms operated or planned by the Midcontinent Independent System Operator (MISO), which delivers safe, cost-effective electric power to 42 million customers across 15 U.S. states and the Canadian province of Manitoba. MISO's grid includes 71,800 miles of transmission lines and a generation capacity of 177,760 MW for a peak summer system demand of 127,125 MW. We collected eight days of location-specific quarter-hourly \emph{wind speed} and \emph{wind direction} values, starting from September 2020, for each of these wind farms. The wind speed is reported in meters per second (m/s), and the wind direction is reported in cardinal (or compass) directions (in degrees) from which it originates. Fig.~\ref{fig:raw-data} presents the spatial map of the wind farms and the wind data at two specific times.


The wind data is prepared at different time and space resolutions. Denote by $K=506$ the total number of wind farms in the region of interest and by $T=750$ the total number of time units (15 minutes per unit) in the time horizon.
The data resolution can be defined as $\xi \coloneqq (\xi^\text{time}, \xi^\text{loc}) \in \{1,2,\dots, T\} \times \{1,2,\dots, K\}$, where $\xi^\text{time}$ and $\xi^\text{loc}$ denote the time and space resolution, respectively. In particular, for the time domain, the time horizon can be evenly divided into $\xi^\text{time}$ frames (Fig.~\ref{fig:map-diff-spatial-resolution}(a)(b)). For the space domain, wind farms can be partitioned into $\xi^\text{loc}$ clusters using the $k$-means algorithm based on their Euclidean distances (Fig.~\ref{fig:map-diff-spatial-resolution}(c)(d)). 
Therefore, a specific spatial cluster at a specific time frame becomes a spatio-temporal subregion. When all these subregions under $i$-th resolution $\xi_i = (\xi^\text{time}, \xi^\text{loc})$ (total of $\xi_i^\text{time} \times \xi^\text{loc}$ subregions) are systematically arranged and indexed, the $j$-th subregion can be represented by the spatio-temporal centroid, denoted as $x_{i, j}=(t_{i, j}, s_{i, j})$, where $t_{i, j}$ and $s_{i,j}$ are the centroid of the associated time frame and spatial cluster, respectively. The wind speed and direction at each subregion $x_{i, j}$ are collected as the average of the observations recorded at the wind farms within the corresponding spatial cluster during the corresponding time frame.

Fig.~\ref{fig:map-diff-spatial-resolution} also highlights that the wind speed changes more frequently than the wind direction, suggesting that the variation of wind speed in this study is highly dynamic. 
A preliminary analysis suggests that the wind data exhibit clear spatio-temporal correlations and decay over time and space, which is illustrated in Fig.~\ref{fig:mae-vs-dist} in Appendix~\ref{app:method-detail}.

\subsection{Spatio-temporal regressive model with lags}
\label{sec:regression}


A broad range of predictive models can be adopted for wind speed prediction, including both physical models \citep{giebel2011state, LEI2009915} and statistical models \citep{erdem2011arma, welch1995introduction}. In this section, we introduce our novel choice of the predictive model $f_i$ for the wind speed under each specified data resolution $\xi_i$. For the notation simplicity we omit the resolution $\xi_i$ in the following discussion since the predictive model under every single resolution is established independently.

The model is motivated by Hawkes process \citep{hawkes1971spectra}, and is carefully designed as a data-driven approach to offer precise wind speed predictions by capturing the spatio-temporal dependencies among wind speed data through a correlation function.
In particular, the wind speed at downstream wind farms can be influenced by those at upstream wind farms due to the wind propagation caused by global pressure gradients \citep{alexiadis1999wind} and the wake effect of wind farms over large spatial scales \citep{platis2018first, porte2020wind}.
Therefore, we can implement wind speed forecasting using historical records of wind speed and direction.
Specifically, the wind speed prediction at $j$-th subregion with centroid $x_{i, j}$ can be specified by:
\begin{equation}
\begin{aligned}
    f_i(x_{i, j}) 
    =&~\nu_i(x_{i, j}) + \sum_{x_{i, j'} \in \mathcal{B}_{x_{i, j}}} g_i(x_{i, j}, x_{i, j'}),
    \quad j \in \mathcal{J}_i.
    \label{eq:wind-speed-predictor}
\end{aligned}
\end{equation}
Here $\nu_i(x_{i, j})$ is a learnable scalar representing the background wind speed at each subregion $x_{i, j}$. The $g_{i}$ is a spatio-temporal correlation function, which characterizes the influence of wind speed at upstream wind farms to downstream wind farms.
Here we assume that such an influence propagates from one place to another along the observed wind direction.
The wind directions, from cluster to cluster, can be specified by a directed \emph{dynamic} graph $G_{i} = (\mathcal{V}_i, \{\mathcal{E}_{i, t_{i, j}}\})$, where $\mathcal{V}_i$ represents the spatial clusters under resolution $\xi_i$, $\mathcal{E}_{i, t_{i, j}} \subseteq \mathcal{E}_i$
is a set of directed edges connecting two clusters if the wind blows from the source cluster to the target cluster at time $t_{i, j}$, and $\mathcal{E}_i$ denotes the edges of the fully-connected graph with vertices $\mathcal{V}_i$. Consequently, we have $g_i(x_{i, j}, x_{i, j'}) \neq 0$ only if $(s_{i,j'}, s_{i,j}) \in \mathcal{E}_{i, t_{i, j'}}$. Besides, we also assume that the influence of wind at upstream wind farms decays quickly over time, thus we only consider the influence of historical observations up to time lag $d$. The spatio-temporal neighbors $\mathcal{B}_{x_{i, j}}$ is finally defined as $\{x_{i, j'}|t_{i, j}-d \leq t_{i, j'} \leq t_{i, j}-1, (s_{i,j'}, s_{i,j}) \in \mathcal{E}_{i, t_{i, j'}}\}$.

The determination of triggering function $g_i$ can be flexible. In our wind example, we define the $g_i$ as:
$   
    g_i(x_{i, j}, x_{i, j'})
    = \alpha_i^{s_{i,j},s_{i,j'}} \beta_i^{s_{i,j'}} \exp\{-\beta_i^{j'} (t_{i, j} - t_{i, j'} - \lambda_{i}^{s_{i,j},s_{i,j'}})\} \cdot y_{i, j'} \cdot \mathbbm{1}\{t_{i, j} - t_{i, j'} \ge \lambda_{i}^{s_{i,j},s_{i,j'}}\},
$
where $\{\lambda_{i}^{s_{i,j},s_{i,j'}} > 0\}$ is a tensor of wind travel times (in seconds) from cluster $s_{i, j'}$ to cluster $s_{i, j}$ at time $t_{i, j'}$ estimated from real data and $\beta_{i}^{s_{i,j'}} \ge 0$ is the decay rate of cluster $s_{i, j'}$'s influence.
The choice of our triggering function $g_i$ can be justified based on three key assumptions.
(1) The upstream influence decays over time and hence the triggering function includes an exponential function $\beta \exp\{-\beta(t-\tau)\}$ that are commonly used to represent such decay, 
where $t$ and $\tau$ are current and historical time, respectively, with $t > \tau$. The parameter $\beta \ge 0$ captures the decay rate of the influence (note that the function integrates to one over $t$).
(2) The inter-cluster influence varies from pair to pair and may depend on the geographical features of the region that lie between two clusters. Hence each edge $(j', j) \in \mathcal{E}_i$ is associated with a non-negative weight $\alpha_{i}^{s_{i,j},s_{i,j'}} \ge 0$ indicating the correlation between cluster $s_{i, j}$ and $s_{i, j'}$: the larger the weight $\alpha_{i}^{s_{i,j},s_{i,j'}}$, the cluster $s_{i, j}$ is more likely to be affected by cluster $s_{i, j'}$.
(3) The upstream influence has a physical propagation delay as the wind must travel over the Earth's surface to reach the downstream cluster \citep{hwang2019do};
such delay can be estimated by the distance between two clusters divided by the wind speed at that time. 
As we can see, the future wind speed can be estimated by the regression of historical observations using practically interpretable coefficients. Thus, we term the proposed $f_i$ as spatio-temporal regressive model with lags.



\subsection{Numerical results}
\label{sec:result}

\begin{figure}[!t]
\centering
\begin{subfigure}[h]{.4\linewidth}
\includegraphics[width=\linewidth]{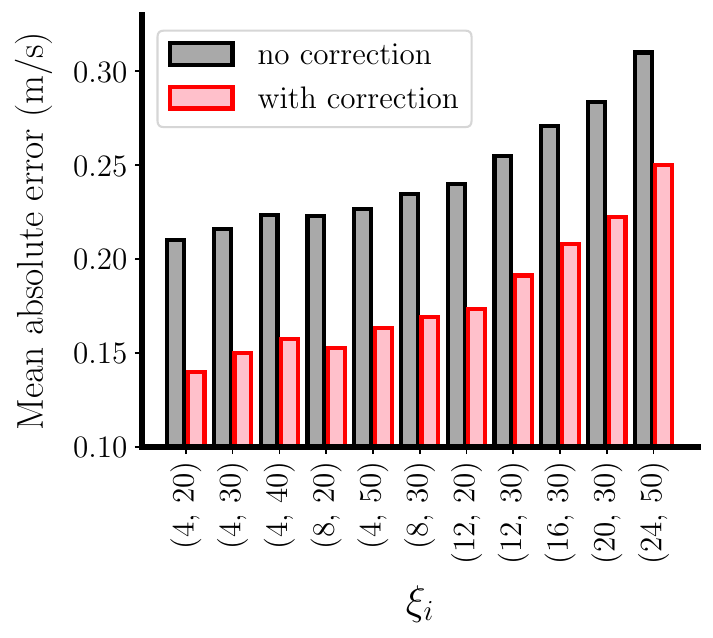}
\caption{Prediction MAE}
\end{subfigure}
\begin{subfigure}[h]{.4\linewidth}
\includegraphics[width=\linewidth]{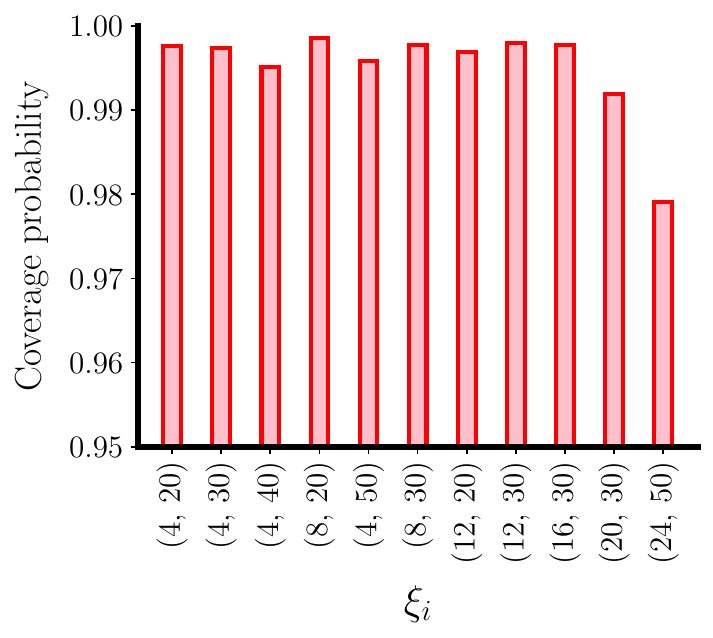}
\caption{Uncertainty interval coverage}
\end{subfigure}
\caption{Performance of our methods on 11 data sets at different data resolutions. The black and red lines represent the MAE of \texttt{STRL} and \texttt{MRSTGP}, respectively. The resolutions are ordered by their product, \ie, $\xi_i^{\text{time}} \times \xi_i^{\text{loc}}$.
}
\label{fig:multi-resolution-comparison}
\end{figure}

After the determination of model details, we present the numerical results of fitting the comprehensive wind data using our proposed methods, which demonstrate the effectiveness of our method and superiority against other baselines in real-world practice.
In our experiment, we test our methods on the wind data with 11 different data resolutions, as shown in Fig.~\ref{fig:multi-resolution-comparison}. 
See more details for data overview in Section~\ref{sec:wind-data-app}. 
To train the spatio-temporal regressive model with lags (\texttt{STRL}), we use Stochastic Gradient Descent (SGD) with a learning rate of $10^{-2}$ and a scaling factor $\delta = 0.8$ to penalize the overestimation of wind speed. Note that the estimated model for predicting wind speed at time $t$ can be used as the warm start for the model at time $t+1$. Details of the training objective of \texttt{STRL} can be found in Appendix~\ref{app:method-detail}.
After fitting the predictive models, we perform the wind speed predictions across the domain under each specified data resolution, and use the observations and predictions to train the multi-resolution spatio-temporal Gaussian process (\texttt{MRSTGP}).
We choose $M = 500$ inducing variables, which is estimated using SGD with a learning rate of $10^{-2}$ and a batch size of $1,000$.
All experiments are performed on Google Colaboratory (Pro version) with 12GB RAM and dual-core Intel processors, with speeds up to 2.3 GHz (without GPU).

\begin{figure}[!t]
\centering
\begin{subfigure}[h]{.7\linewidth}
\includegraphics[width=.48\linewidth]{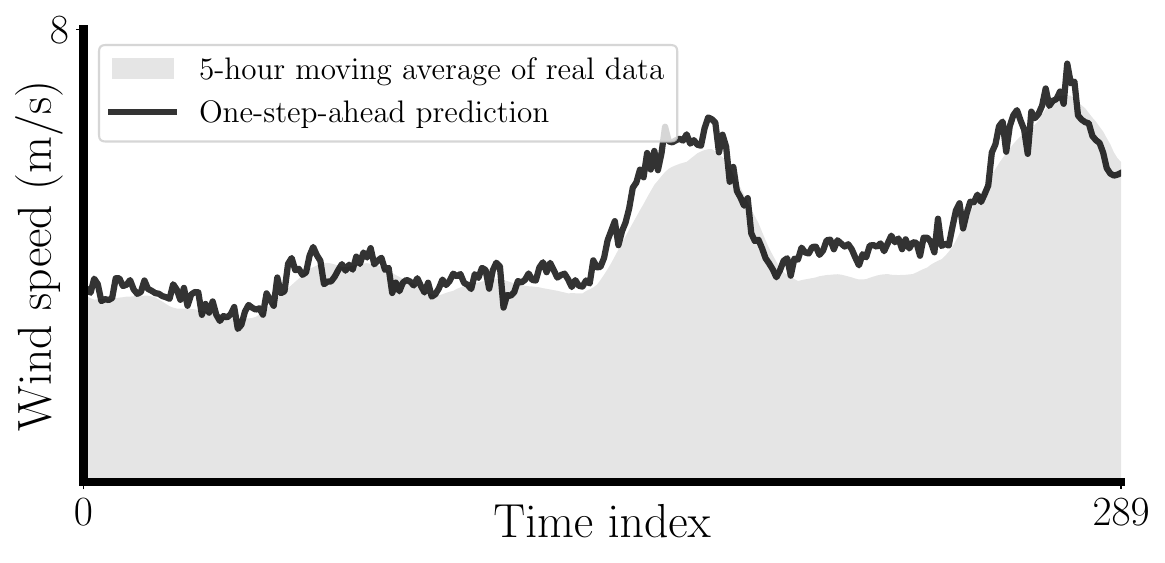}
\includegraphics[width=.48\linewidth]{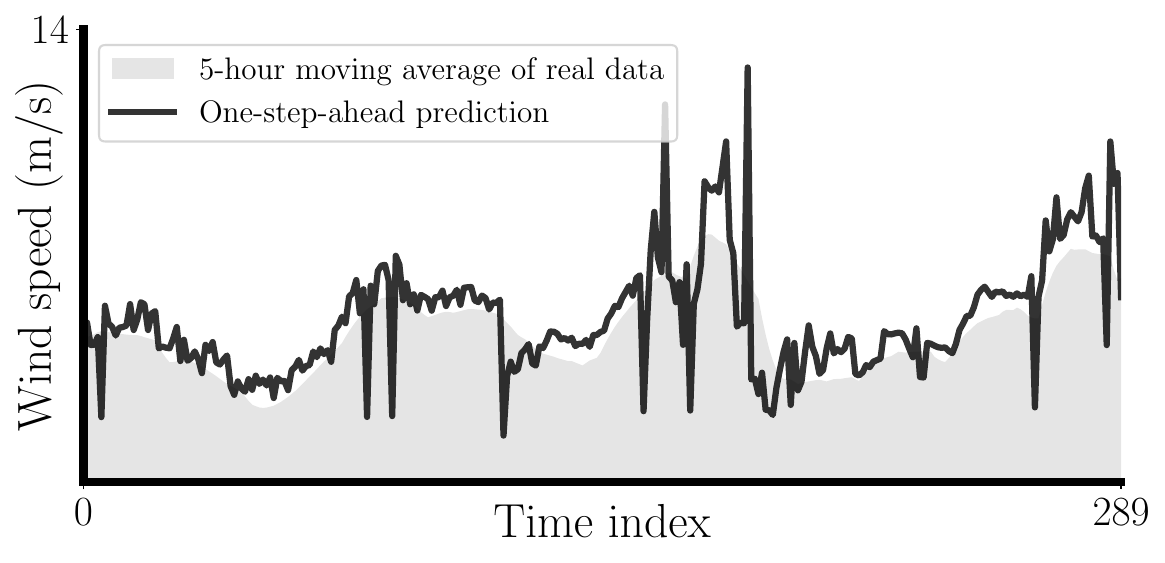}
\caption{$\xi=(12,20)$}
\end{subfigure}
\vfill
\begin{subfigure}[h]{.7\linewidth}
\includegraphics[width=.48\linewidth]{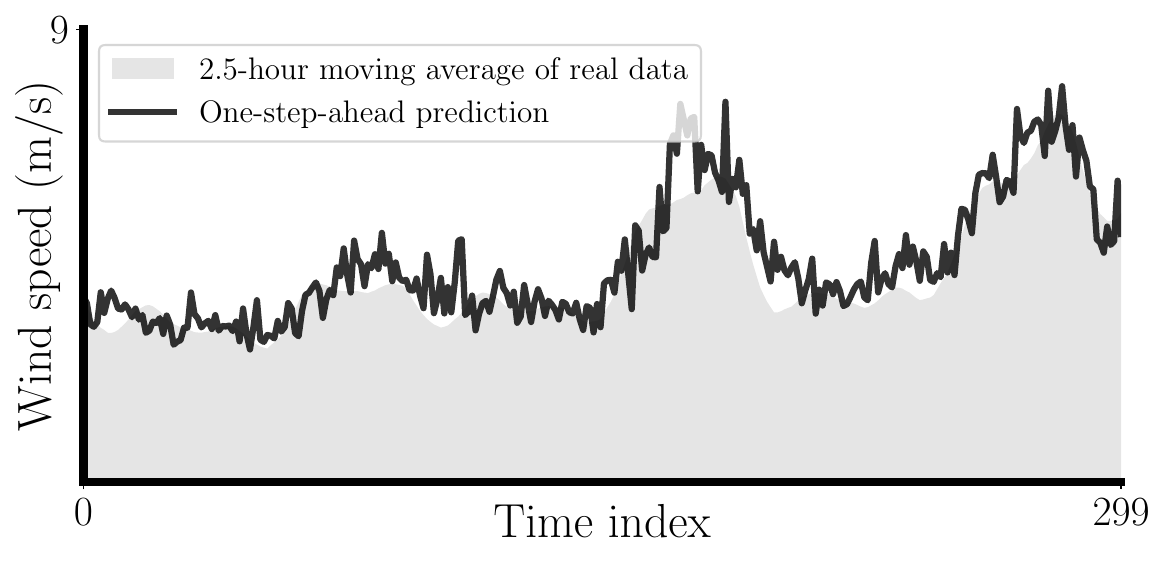}
\includegraphics[width=.48\linewidth]{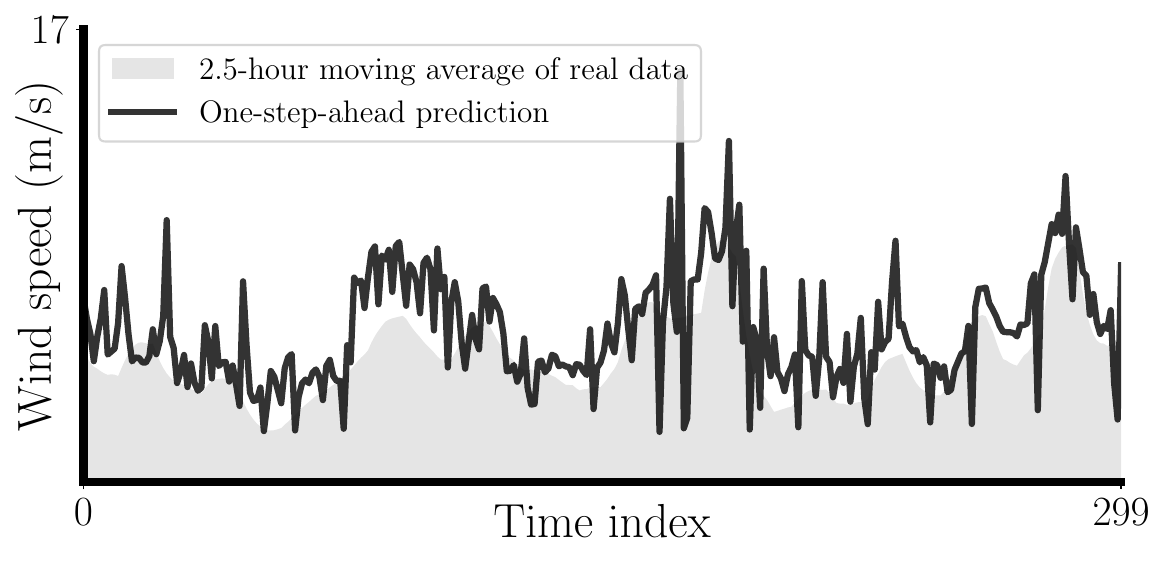}
\caption{$\xi=(24,20)$}
\end{subfigure}
\vfill
\begin{subfigure}[h]{.7\linewidth}
\includegraphics[width=.48\linewidth]{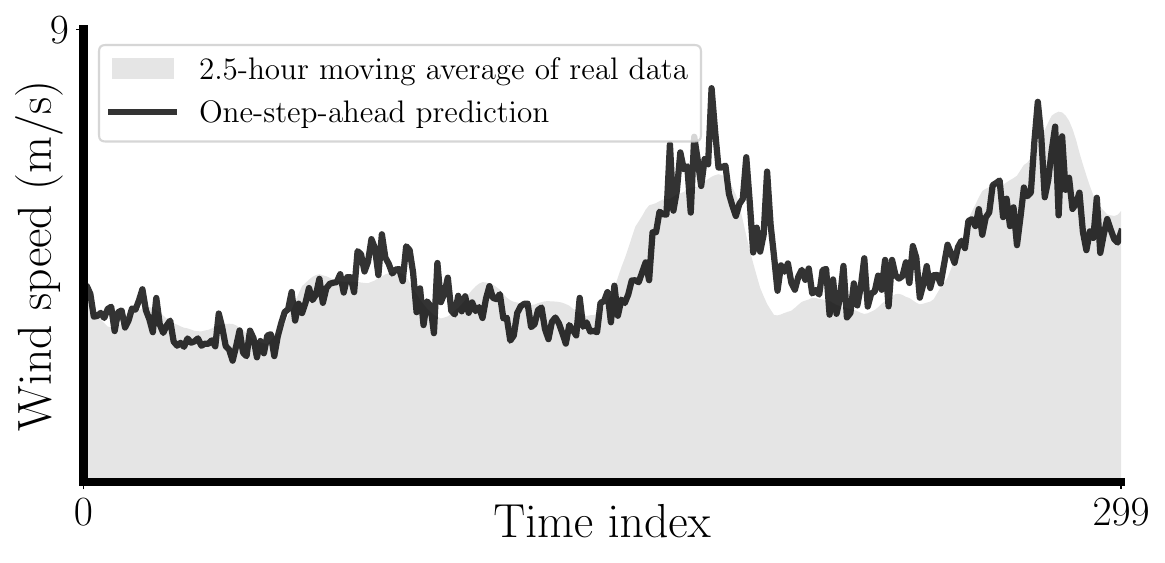}
\includegraphics[width=.48\linewidth]{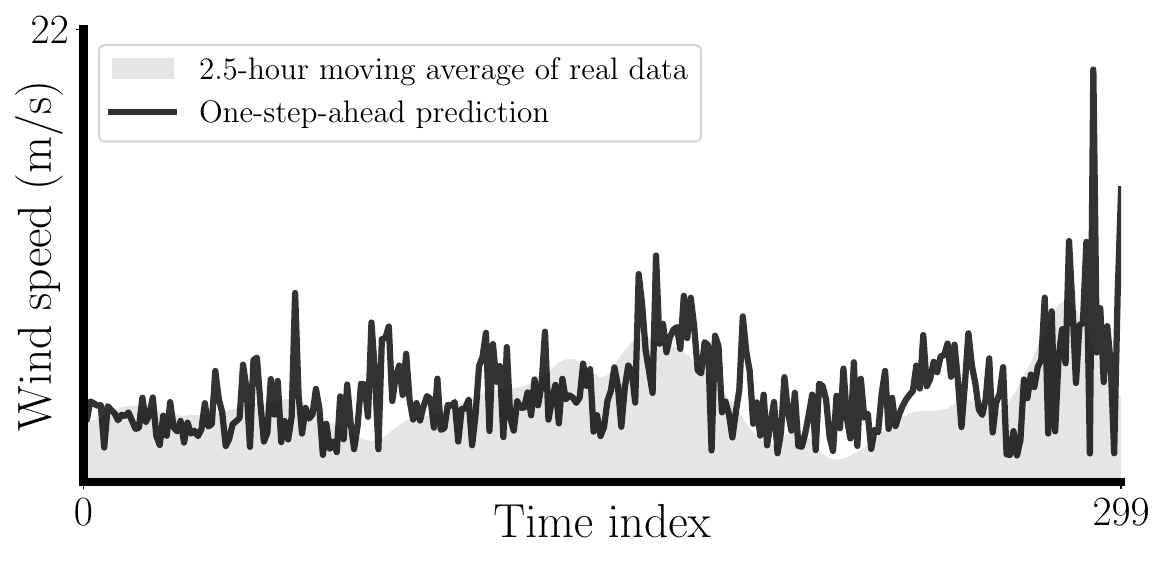}
\caption{$\xi=(24,50)$}
\end{subfigure}
\caption{Examples of wind speed predictions using \texttt{STRL} at different data resolutions. The first column shows the spatial average of prediction and the second column shows the prediction of one cluster. The grey-shaded areas represent the ground truth.
}
\label{fig:pred-out-of-sample}
\end{figure}

We elaborate our training and predicting procedures as following: for any given time frame $t$: (i) We first train \texttt{STRL} at each resolution $\xi_i$ using training data $\{(x_{i,j},y_{i,j})\}_{t_{i,j} \leq t}$ up to time $t$. Then we use the learned \texttt{STRL}s to make in-sample one-step-ahead prediction $\{\widehat f_i(x_{i,j})\}_{t_{i,j}\leq t}$ and obtain the prediction errors $\{\epsilon_{i,j}\}_{t_{i,j}\leq t}$ by \eqref{eq:proposed-framework-of-variable}. 
(ii) We then fit the \texttt{MRSTGP} using prediction errors up to time $t$.
(iii) Finally, we make out-of-sample prediction $\{\widehat{f}_i(x_{i,j})\}_{t_{i,j}=t+1}$ at time $t+1$ by feeding the data from $t-d$ to $t$ into the fitted \texttt{STRL}s. In the meantime, we estimate the prediction errors at time $t+1$ using the fitted \texttt{MRSTGP} (see \eqref{eq:pred-posterior}). Thus, the corrected predictions of wind speed at time $t+1$ can be obtained by adding the estimated prediction errors to the original out-of-sample predictions.
To evaluate the model's predictive performance, we measure the mean absolute error (MAE) of one-step-ahead out-of-sample wind speed prediction for \texttt{STRL}, \texttt{MRSTGP} and four baselines over time horizon $[240, 550]$. The data before $t=240$ are always used for model training, and the out-of-sample predictions are performed at any time during $[240, 550]$.

\paragraph{Spatio-temporal prediction}

The \texttt{STRL}’s predictive power is assessed by performing the one-step-ahead (out-of-sample) prediction at different data resolutions. 
The prediction for time index $t$ given resolution $\xi_i$ is carried out by 
(i) withholding the data after $t$ from the model estimation and using the $\xi_i^{\text{time}}$-unit moving average of historical data before $t$ to fit the model; and 
(ii) using the fitted model to make predictions for the (hold-out) data at time $t+1$.
Fig.~\ref{fig:pred-out-of-sample} presents examples of one-step-ahead predictions of \texttt{STRL} at three different data resolutions over time.
We note that the out-of-sample prediction results in Fig.~\ref{fig:pred-out-of-sample} are carried out at more than 280 time steps.
The above results show that model \texttt{STRL} can predict wind speed accurately at the cluster level. 
Observe that high-resolution wind speeds oscillate rapidly, with significant amplitude over time and space, and increases in data resolution $\xi_i$ will degrade the predictive accuracy significantly. 

\paragraph{Corrected prediction for multi-resolution data} 

\begin{figure}[!t]
\centering
\begin{subfigure}[h]{.35\linewidth}
\includegraphics[width=\linewidth]{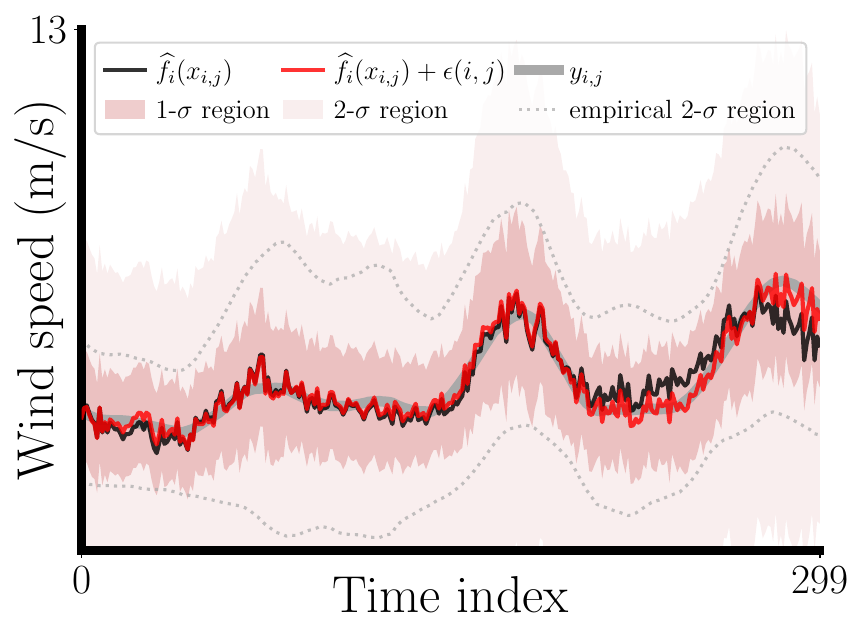}
\caption{$\xi=(24,50)$}
\end{subfigure}
\begin{subfigure}[h]{.35\linewidth}
\includegraphics[width=\linewidth]{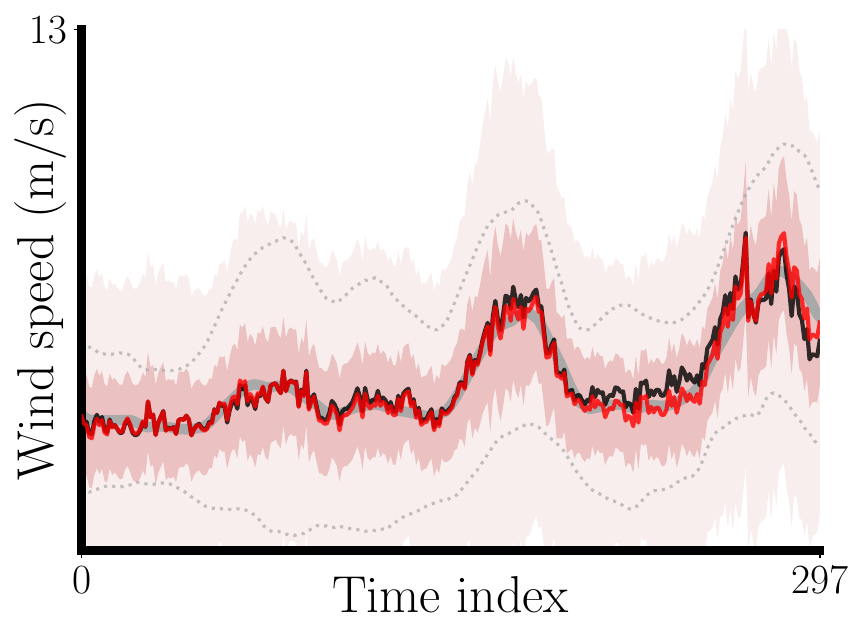}
\caption{$\xi=(20,30)$}
\end{subfigure}
\vfill
\begin{subfigure}[h]{.35\linewidth}
\includegraphics[width=\linewidth]{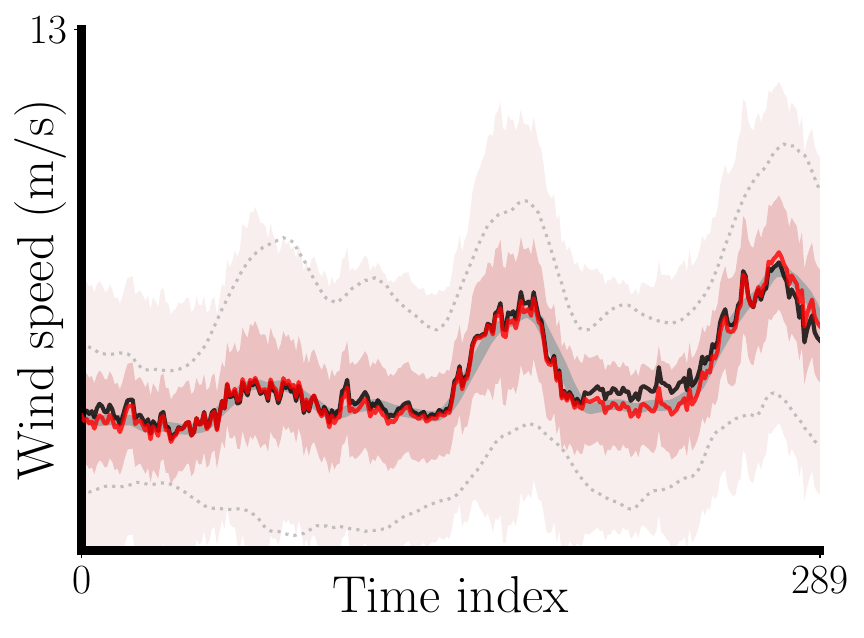}
\caption{$\xi=(12,30)$}
\end{subfigure}
\begin{subfigure}[h]{.35\linewidth}
\includegraphics[width=\linewidth]{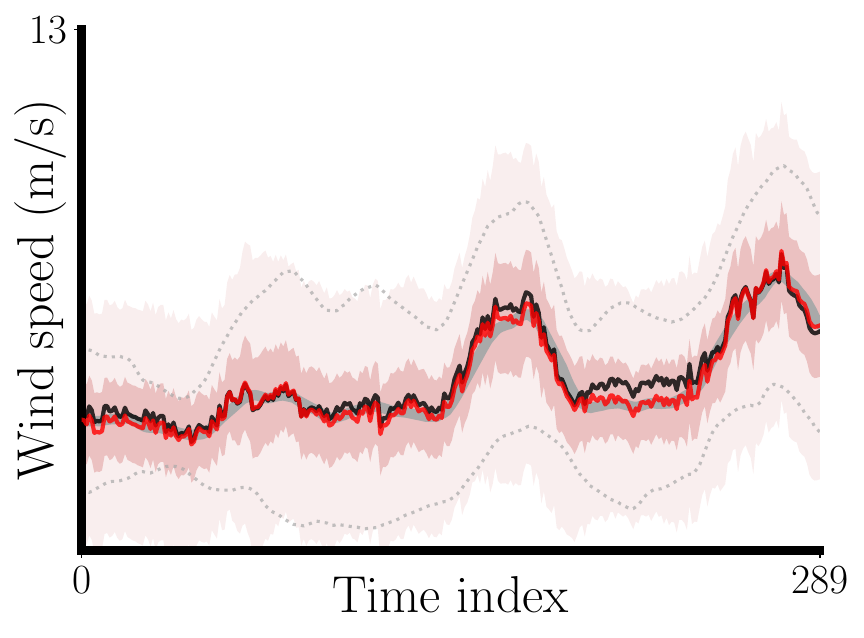}
\caption{$\xi=(12,20)$}
\end{subfigure}
\caption{
Examples of the effectiveness of prediction correction by \texttt{MRSTGP} on data sets at different data resolutions. The grey and black lines represent the spatially averaged observations and predictions by \texttt{STRL}, respectively. The red lines represent the spatially averaged corrected predictions by \texttt{MRSTGP}, with the shaded areas representing the corresponding confidence interval of the predictions. 
}
\label{fig:ci-out-of-sample}
\end{figure}

\begin{figure}[!t]
\centering
\begin{subfigure}[h]{.22\linewidth}
\includegraphics[width=\linewidth]{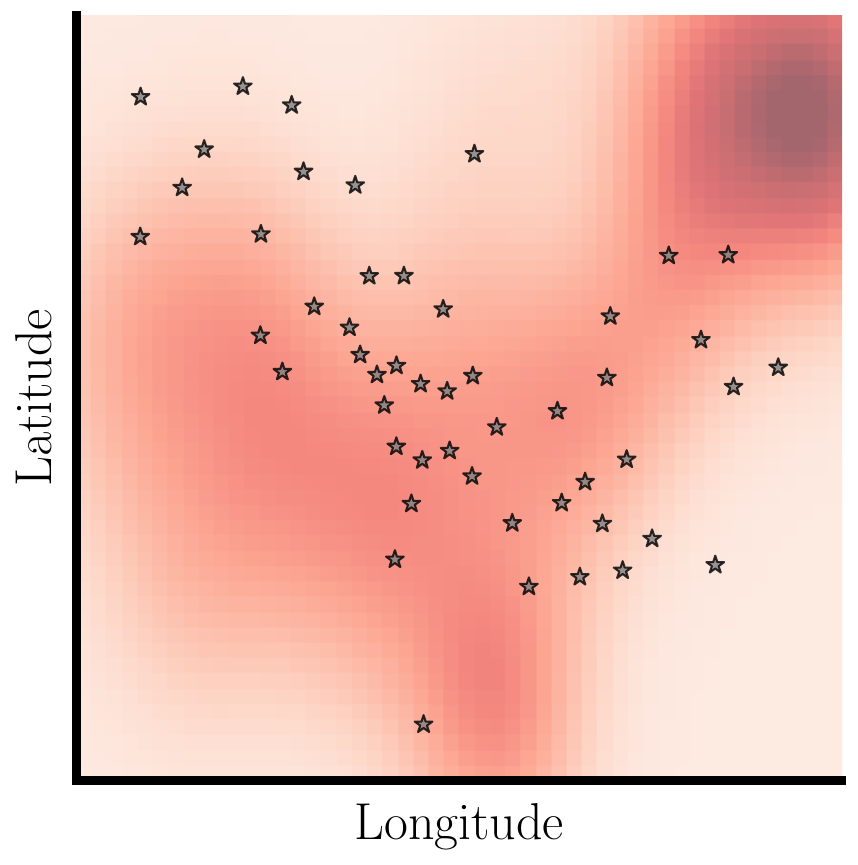}
\caption{$\xi=(24,50)$}
\end{subfigure}
\begin{subfigure}[h]{.22\linewidth}
\includegraphics[width=\linewidth]{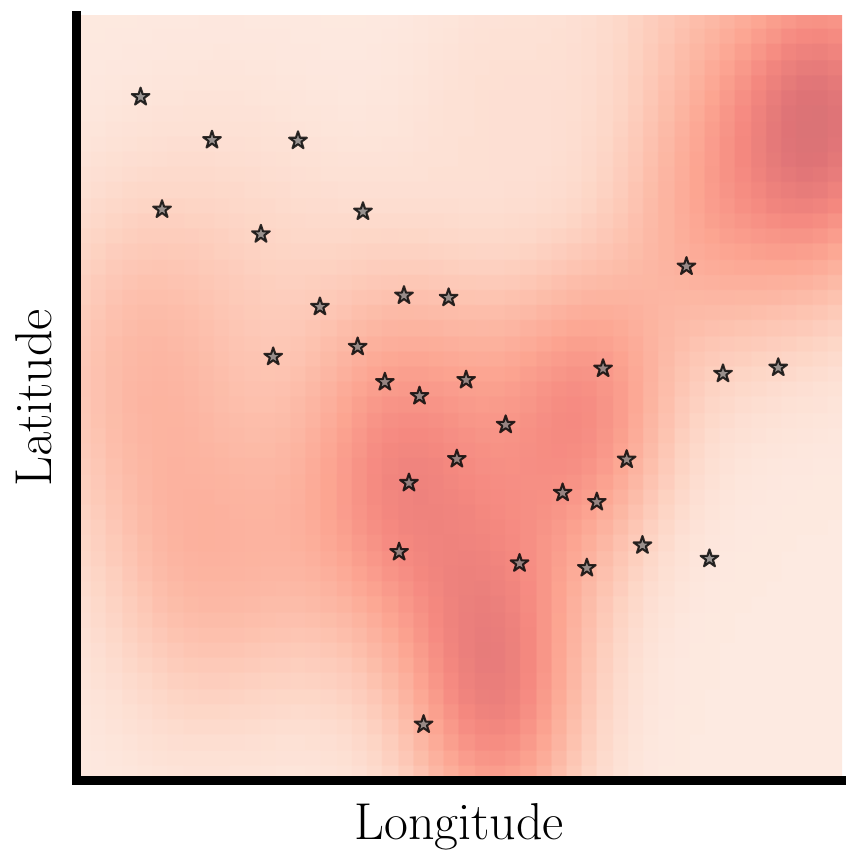}
\caption{$\xi=(20,30)$}
\end{subfigure}
\begin{subfigure}[h]{.22\linewidth}
\includegraphics[width=\linewidth]{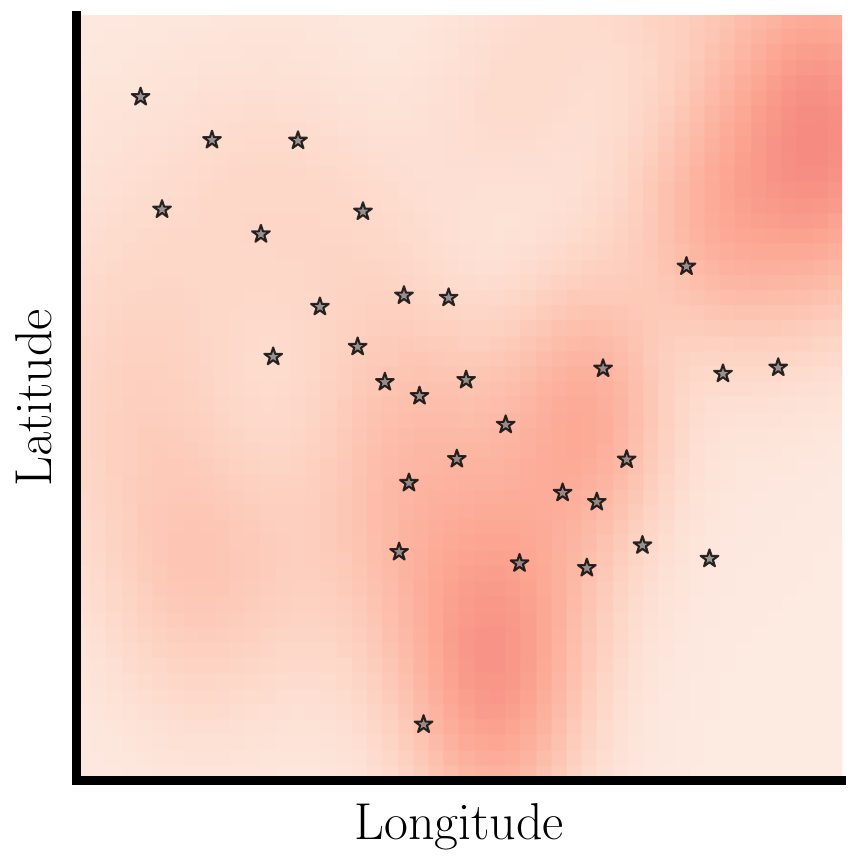}
\caption{$\xi=(12,30)$}
\end{subfigure}
\begin{subfigure}[h]{.22\linewidth}
\includegraphics[width=\linewidth]{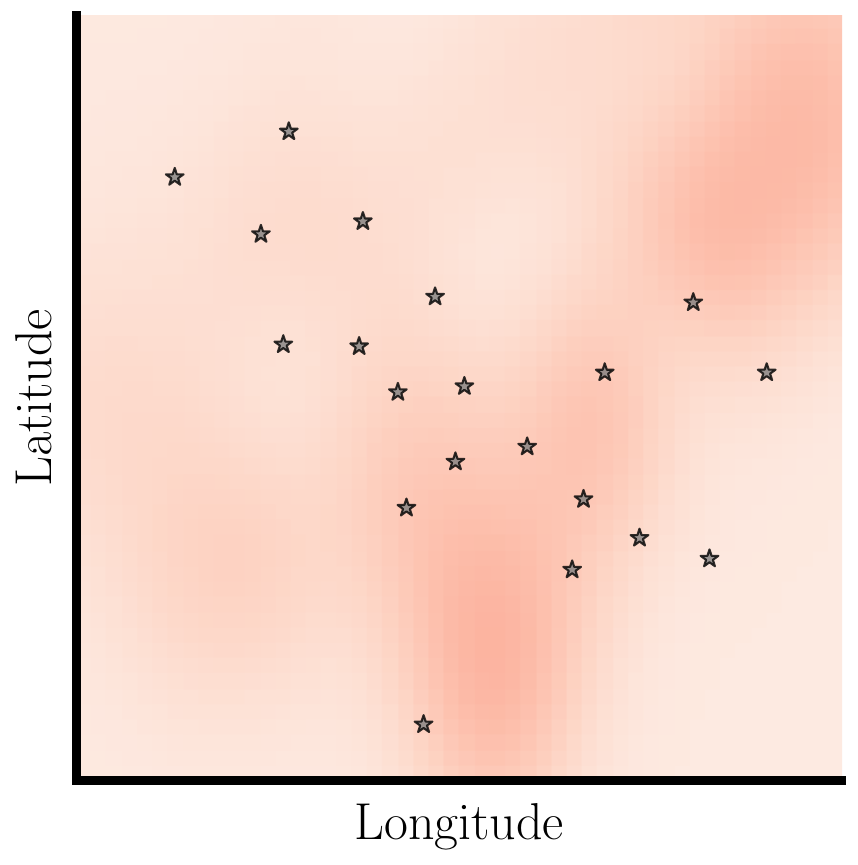}
\caption{$\xi=(12,20)$}
\end{subfigure}
\begin{subfigure}[h]{.035\linewidth}
\includegraphics[width=\linewidth]{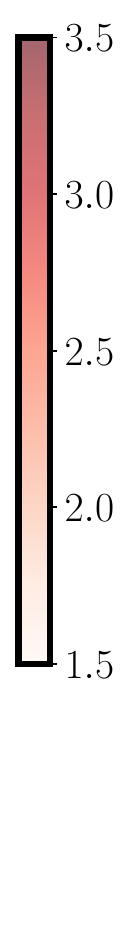}
\end{subfigure}
\caption{Examples of the spatial distribution of $1\sigma$ confidence interval (68.27\%) for the corrected prediction suggested by \texttt{MRSTGP} on four data sets at different data resolutions. The star markers represents the centroid of cluster; the color depth represents the width of confidence interval (m/s).}
\label{fig:ci-out-of-sample-space}
\end{figure}

The learning of model \texttt{MRSTGP} uses all the data sets at different resolutions and their predictions before time index $t$ as input to correct the wind speed prediction at the next time step $t+1$. Fig.~\ref{fig:ci-out-of-sample} shows four examples of corrected predictions on four data sets with different resolutions. 
The blue and red lines indicate the predictions of \texttt{STRL} and \texttt{MRSTGP}, respectively. 
The results show that the \texttt{MRSTGP} outperforms the \texttt{STRL}, particularly for the period from 180 to 250.

Fig.~\ref{fig:ci-out-of-sample} also presents the estimated confidence interval suggested by \texttt{MRSTGP}. The black dots represent the observed wind speeds for a specific cluster; the light and dark shaded blue areas indicate the 1-$\sigma$ (68.27\%) and 2-$\sigma$ (95.45\%) prediction intervals, respectively. The result shows that the estimated confidence interval achieves good data coverage. Fig.~\ref{fig:ci-out-of-sample-space} presents the spatial distribution of the estimated 1-$\sigma$ (68.27\%) prediction intervals for the same four data sets. The model achieves a smaller confidence interval as the data resolution decreases. Note also that the regions with significantly higher predictive uncertainty are where the wind usually originated because the data sets do not have enough upstream observations to infer the wind condition for these regions. Fig.~\ref{fig:multi-resolution-comparison} compares the predictive performance between \texttt{STRL} and \texttt{MRSTGP} quantitatively on each of the data sets. The results show that the MAEs of \texttt{MRSTGP} are significantly lower than those of \texttt{STRL} for all scenarios and confirm the effectiveness of \texttt{MRSTGP}.

\paragraph{Comparison with baselines}

\begin{table}[!t]
\caption{Performance comparison of all methods. The prediction MAE (lower is better) of all models are reported by averaging over all resolutions (the first column) and at two randomly selected resolutions (the last two columns), respectively.}
\label{tab:mae}
\centering
\resizebox{.9\linewidth}{!}{%
    \begin{tabular}{cccc}
    \hline
    \hline
    & MAE (average) & MAE ($\xi_i=(24, 50)$) & MAE ($\xi_i=(4, 20)$)  \\
    \hline
    \texttt{NN} & $1.48 (41\%)$ & $1.37 (38\%)$ & $1.45 (41\%)$\\
    \texttt{MR-NN} & $0.79 (20\%)$ & $0.67 (16\%)$ & $0.73 (19\%)$\\
    \texttt{LSTM} & $1.44 (37\%)$ & $1.78 (46\%)$ & $1.16 (29\%)$\\
    \texttt{MR-LSTM} & $1.30 (34\%)$ & $1.32 (33\%)$ & $1.04 (28\%)$\\
    \texttt{VAR} & $1.28 (34\%)$ & $1.26 (31\%)$ & $0.52 (14\%)$\\
    \texttt{MR-VAR} & $1.06 (29\%)$ & $0.79 (19\%)$ & $0.43 (12\%)$\\
    \hline
    \texttt{DeepTCN} & $0.53 (13\%)$ & $0.76 (17\%)$ & $0.27 (6\%)$\\
    \texttt{DeepAR} & $0.55 (13\%)$ & $0.74 (16\%)$ & $0.26 (6\%)$\\
    \hline
    \texttt{STRL} & $0.27 (6\%)$ & $0.31 (7\%)$ & $0.21 (6\%)$\\
    \texttt{MRSTGP} & $\textbf{0.20(5\%)}$ & $\textbf{0.25(5\%)}$ & $\textbf{0.14(4\%)}$\\
    \hline
    \hline
    \end{tabular}
    %
}
\end{table}

We compare the proposed models with the following four state-of-the-art baselines (See \cite{LEI2009915, soman2010review} for detailed reviews): (i) Neural network-based univariate prediction model (\texttt{NN}) \citep{CATALAO20111245} (ii) Vector autoregressive (\texttt{VAR}) \citep{lutkepohl2013vector} (iii) Long Short-Term Memory (\texttt{LSTM}) \citep{hochreiter1997long} (iv) Global deep learning multivariate probabilistic forecasting models, including \texttt{DeepAR} model \citep{SALINAS20201181} and \texttt{DeepTCN} model \citep{CHEN2020491}. Please refer to Appendix~\ref{append:add-exp} for detailed descriptions of baselines.
Table~\ref{tab:mae} reports the average MAE (and the percentage to the ground truth) of out-of-sample prediction given by our approaches and baselines.  
The results confirm that the proposed models significantly outperform the baseline methods by providing forecasts more accurately. 

Besides, we integrate our multi-resolution correcting framework with the first three baselines, yielding \texttt{MR-NN}, \texttt{MR-LSTM}, and \texttt{MR-VAR} models. The purpose of this ablation study is to demonstrate the effectiveness of our framework that leverages information from multi-resolution data sources for prediction enhancement. The prediction errors of all three baselines on testing data are reduced after applying the multi-resolution GP framework. Meanwhile, \texttt{STRL} consistently outperforms the baselines across different resolutions, both before or after the model correction. These results highlight the superiority of our proposed \texttt{STRL} in wind predictions against baselines with the capability to capture complex spatio-temporal dependencies among wind data.

\section{Discussion}
This paper presents a novel spatio-temporal predictive model for wind speed that incorporates a directed dynamic graph to represent the wind directions between clusters of wind farms. Additionally, a Bayesian framework is introduced to bridge the gap between data at different resolutions through a Gaussian process, thereby enhancing the model's predictive power. The joint framework exhibits promising results in modeling and predicting wind speed, as demonstrated by a numerical study. The proposed methods outperform existing approaches in terms of predictive performance and provide reasonable uncertainty quantification.


\bibliographystyle{apalike}
\bibliography{arxiv_refs}

\newpage
\appendix

\section{Extracting DDG}
\label{append:ddg}

\begin{figure}[!h]
\centering
\begin{subfigure}[h]{0.3\linewidth}
\includegraphics[width=\linewidth]{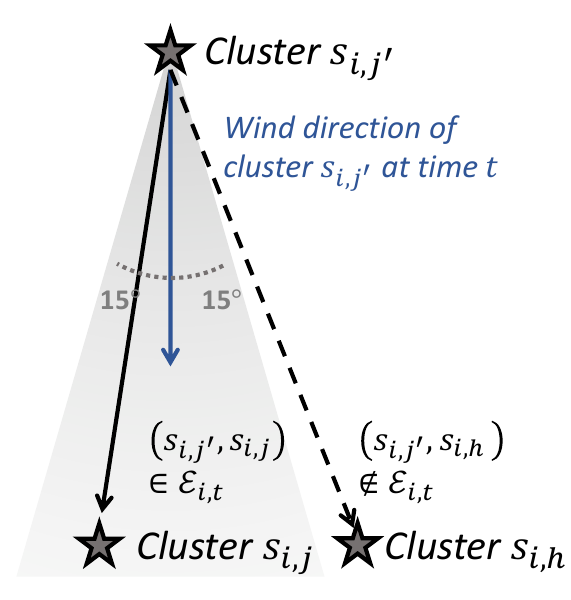}
\caption{Dynamic edge}
\end{subfigure}
\begin{subfigure}[h]{0.3\linewidth}
\includegraphics[width=\linewidth]{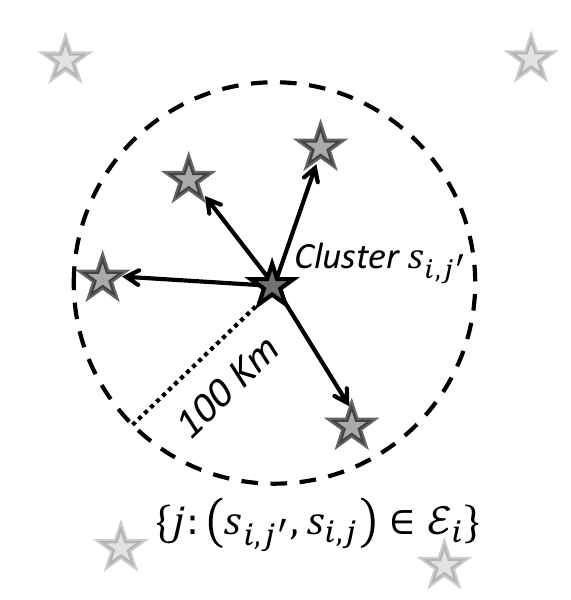}
\caption{Graph support}
\end{subfigure}
\caption{Illustrations of (a) an edge in the DDG and (b) the support of DDG for one cluster.}
\label{fig:illustration-ddg}
\end{figure}

\vspace{.1in}
\noindent\emph{Directed dynamic graph for wind direction.}
The directed dynamic graph (DDG) $G_{i} = (\mathcal{V}_i, \{\mathcal{E}_{i, t_{i, j}}\})$ in \eqref{eq:wind-speed-predictor} defines the wind direction between clusters, which can be extracted from the raw data.
Given a data resolution $\xi_i$, the dynamic graph includes the directed edge from cluster $s_{i,j'}$ to cluster $s_{i,j}$ at time $t_{i,j}$ if the difference between their wind directions at that time is not larger than 15\degree, as illustrated in Fig.~\ref{fig:illustration-ddg} (a). The graph support $\mathcal{E}_i$ has a sparse structure: indeed, the preliminary analysis in Section~\ref{sec:wind-data-app} indicates that an arbitrary cluster can only be affected by its nearest clusters within a 100-$Km$ radius of itself, as shown in Fig.~\ref{fig:illustration-ddg} (b). The sparsity of the graph support leads to significant reductions in the calculation of \eqref{eq:wind-speed-predictor} and plays a big role in the computational efficiency of the proposed model. 

\section{Methodology detail}
\label{app:method-detail}

\begin{figure}[!t]
\centering
\begin{subfigure}[h]{0.24\linewidth}
\includegraphics[width=\linewidth]{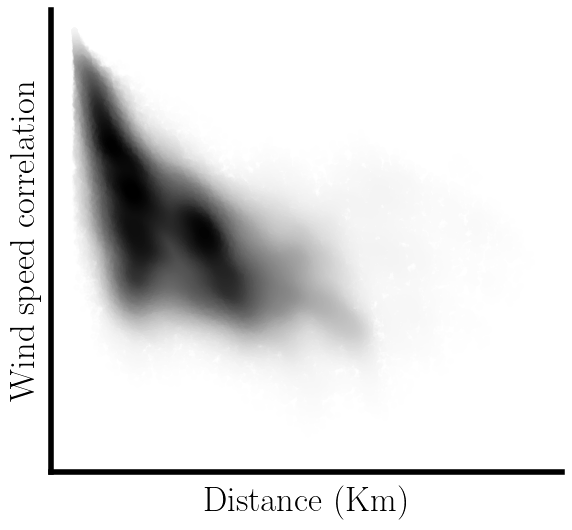}
\caption{}
\end{subfigure}
\begin{subfigure}[h]{0.24\linewidth}
\includegraphics[width=\linewidth]{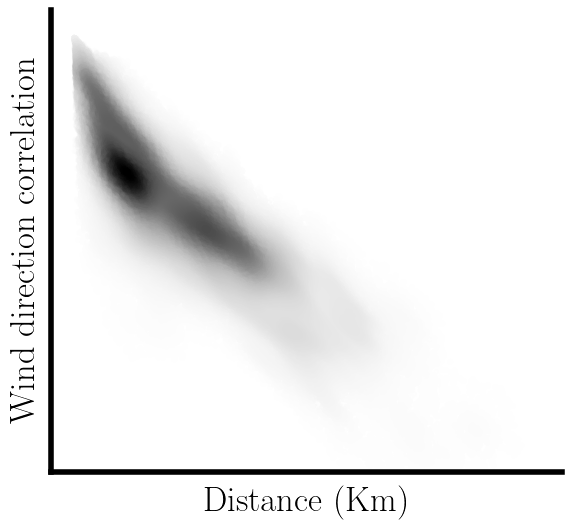}
\caption{}
\end{subfigure}
\begin{subfigure}[h]{0.24\linewidth}
\includegraphics[width=\linewidth]{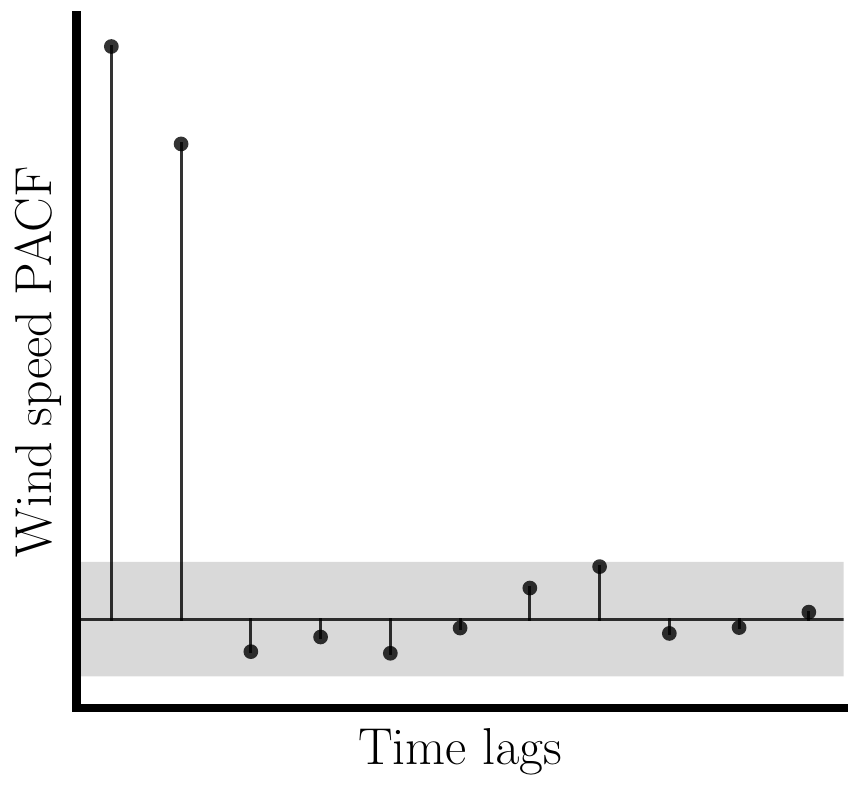}
\caption{}
\end{subfigure}
\begin{subfigure}[h]{0.24\linewidth}
\includegraphics[width=\linewidth]{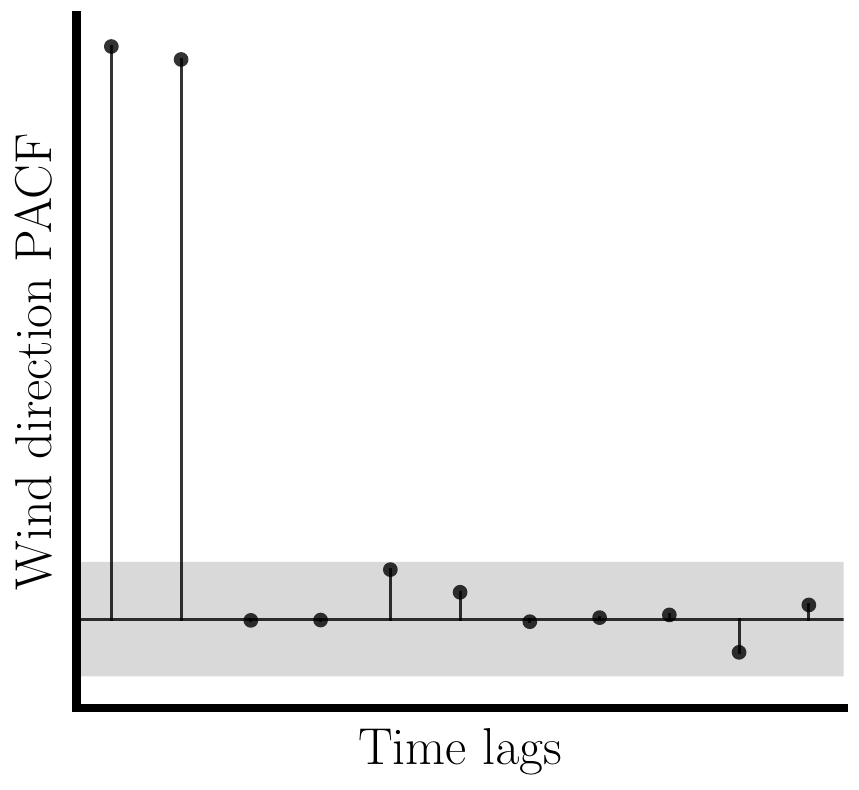}
\caption{}
\end{subfigure}
\caption{Data similarity versus distance and time differences. 
(a)(b) the Pearson correlation coefficients of wind speed and direction between two arbitrary wind farms versus their distance, respectively. The color depth indicates the density of these points; the darker the denser. (c)(d) the partial autocorrelation functions of wind speed and direction at one given wind farm. The shaded areas indicate insignificant values.}
\label{fig:mae-vs-dist}
\end{figure}

\vspace{.1in}
\noindent\emph{Model estimation.}
The model with resolution $\xi_i$ can be estimated by minimizing the mean square error between the true wind speed and its prediction. 
Denote the set of the parameters of the model $f_i$ as $\theta_i \coloneqq \{\{\nu_i\}, \{\alpha^{s_{i,j},s_{i,j'}}_i\}, \{\beta_i^{s_{i,j'}}\}\} \in \Theta_i$, where $\Theta_i \subseteq \mathbb{R}_+^{\xi_i^{\text{loc}}} \times \mathbb{R}_+^{\xi_i^{\text{loc}} \times \xi_i^{\text{loc}}} \times \mathbb{R}_+^{\xi_i^{\text{loc}}}$ is the corresponding parameter space.
It is noteworthy that the model includes less than $\xi_i^{\text{loc}} (2 + \xi_i^{\text{loc}})$ parameters thanks to the sparse structure of $\{\alpha^{s_{i,j},s_{i,j'}}_i\}$. The model can be effectively learned by using $\xi_i^{\text{loc}} \times (T-\xi_i^{\text{time}})$ points under resolution $\xi_i$, where $\xi_i^{\text{loc}} \ll T-\xi_i^{\text{time}}$; for example, when $\xi_i=(4,20)$, there are about $120$ parameters and $3,740$ data points. 
The over-prediction of future wind power may result in fewer commitments of other types of generators, creating reliability issues. To this end, the loss function for training the model includes a scaling factor that penalizes overestimations. 
Formally, the optimal parameters can be found by solving the following optimization problem:
\begin{align*}
&~\underset{\theta_i \in \Theta_i}{\arg \min}~l(\theta_i) \coloneqq \sum_{j\in \mathcal{J}_i}
\Big( \epsilon^2((i, j)) + \delta \mathbbm{1}\left\{y_{i,j} \le \widehat{f}_{i,j}\right\} \epsilon^2((i, j)) \Big) ,
\end{align*}
where 
$\epsilon^2((i, j))$ is the square error of the prediction and $\delta \ge 0$ is a factor that controls the magnitude of penalization for the overestimation. 

\section{Derivation of ELBO}
\label{append:elbo}

Assume the posterior distribution  $p(\epsilon, z|y)$ over random variable vector $\epsilon$ and $z$ is approximated by a variational distribution $q(\epsilon, z)$. 
Suppose this variational distribution $q(\epsilon, z)$ can be factorized as $q(\epsilon, z) = p(\epsilon | z) q(z)$. Hence, the ELBO can be derived as follows:
\begingroup
\allowdisplaybreaks
\begin{align*}
    \log p(y)
    =&~ \log \int \int p(y|\epsilon, z) p(\epsilon, z) d \epsilon d z\\
    =&~ \log \int \int p(y|\epsilon) p(\epsilon, z) d \epsilon d z\\
    =&~ \log \int \int p(y|\epsilon) p(\epsilon, z) \frac{q(\epsilon, z)}{q(\epsilon, z)} d \epsilon d z\\
    =&~ \log \mathbb{E}_{q(\epsilon, z)} \left [  p(y|\epsilon) \frac{p(\epsilon, z)}{q(\epsilon, z)} \right ]\\
    \overset{(i)}{\ge} &~ \mathbb{E}_{q(\epsilon, z)} \log \left [  p(y|\epsilon) \frac{p(\epsilon, z)}{q(\epsilon, z)} \right ]\\
    =&~ \int \int \log p(y|\epsilon) q(\epsilon, z) d \epsilon d z - \int \int \log \left ( \frac{q(\epsilon, z)}{p(\epsilon, z)} \right ) q(\epsilon, z) d \epsilon d z\\
    \overset{(ii)}{=}&~ \mathbb{E}_{q(\epsilon)}\left [ \log p(y|\epsilon) \right ] - \text{KL}\left [ q(z) || p(z) \right ],
\end{align*}
\endgroup
where $q(\epsilon)$ is the marginal of $\epsilon$ from the joint variational distribution $q(\epsilon, z)$, by integrating $z$ out. The inequality $(i)$ holds due to the the Jensen's inequality. The equality $(ii)$ holds because
\begingroup
\allowdisplaybreaks
\begin{align*}
&~ \int \int \log \left ( \frac{q(\epsilon, z)}{p(\epsilon, z)} \right ) q(\epsilon, z) d \epsilon d z\\
= &~ \int \int \log \left ( \frac{p(\epsilon| z) q(z)}{p(\epsilon |z) p(z)} \right ) q(\epsilon, z) d \epsilon d z\\
= &~ \int \int \log \left (\frac{q(z)}{p(z)} \right ) q(\epsilon, z) d \epsilon d z\\
= &~ \int \log \left (\frac{q(z)}{p(z)} \right ) \left ( \int q(\epsilon, z) d \epsilon \right) d z\\
= &~ \int \log \left (\frac{q(z)}{p(z)} \right ) q(z) d z\\
= &~ \text{KL}\left [ q(z) || p(z) \right ].
\end{align*}
\endgroup

To calculate the ELBO, we also need to derive the analytical expression for $q(\epsilon)$ and $\text{KL}[q(z)||p(\epsilon)]$. 
First, given the joint distribution defined in \eqref{eq:joint-dist-f-u}, we apply the multivariate Guassian conditional rule and have the closed-form expression for the conditional distribution:
$
    p(\epsilon|z) = \mathcal{N}(\epsilon|A z, B),
$
where $A = K_{UV} K_{VV}^{-1}$ and $B = K_{UU} - K_{UV} K_{VV}^{-1} K_{UV}^\top$.
Now, due to the factorization $q(\epsilon, z) = p(\epsilon | z) q(z)$, we have
\begingroup
\allowdisplaybreaks
\begin{align*}
    q(\epsilon)
    =&~\int q(\epsilon, z) dz\\
    =&~\int p(\epsilon | z) q(z) dz\\
    =&~\int \mathcal{N}(\epsilon|Az, B)\cdot \mathcal{N}(z|m,S) dz\\
    =&~\int \mathcal{N}(\epsilon|Am, ASA^\top + B)\cdot \mathcal{N}(z|m,S) dz\\
    =&~\mathcal{N}(\epsilon|Am, ASA^\top + B) \cdot \int \mathcal{N}(z|m,S) dz\\
    =&~\mathcal{N}(\epsilon|Am, K_{UU} + A (S - K_{VV}) A^\top).
\end{align*}
\endgroup
Using the fact that both $q(z) = \mathcal{N}(z|m, S)$ and $p(z) = \mathcal{N}(z|\mathbf{0},K_{VV})$ are multivariate Gaussian distributions, the analytical expression for the KL divergence between $q(z)$ and $p(z)$ can be derived as: 
\begin{align*}
    \text{KL}\left [ q(z) || p(z) \right ]
    =&~\frac{1}{2}\bigg( \log\left( \frac{\det(K_{VV})}{\det(S)} \right) - M + \text{tr}(K_{VV}^{-1} S) + (\mathbf{0} - m)^\top K_{VV}^{-1} (\mathbf{0} - m)\bigg),
\end{align*}
where $\det(\cdot)$ is the matrix determinant and $\text{tr}(\cdot)$ is the trace of matrix. 

\section{Stochastic gradient based optimization}
\label{append:sgd}

In this section, we describe our learning algorithm. The optimal parameters of the proposed multi-resolution model can be found by maximizing \eqref{eq:objective} using gradient-based optimization.
However, the full gradient evaluation can still be expensive to be carried out. 
With a sparse prior (inducing variables), even though we can tackle the computational challenge in inverting a big matrix, evaluating the gradient of the first term in \eqref{eq:obj-elbo} still requires the full data set, which is memory-intensive if the size of the data set $N$ is too large.
To alleviate the problem of expensive gradient evaluation, we adopt a stochastic gradient based method \citep{hensman2013gaussian} and only compute the gradient of the objective function evaluated on a random subset of the data at each iteration. 

Specifically, to apply stochastic gradient descent to the variational parameters ($V$, $m$, $S$) in our GP model, we follow the idea of \cite{hensman2013gaussian, martens2014new} by taking steps in the direction of the approximate natural gradient, which is given by the usual gradient re-scaled by the inverse Fisher information. 
The learning algorithm has been summarized in Algorithm~\ref{algo:learning}.

\begin{algorithm}[!ht]
\begin{algorithmic}
    \STATE {\bfseries Initialization:} Randomly initialize $\theta, V, m, S$\;
    \STATE {\bfseries Input:} Data set $U, y, f$; Number of iterations $B$; Batch size $n$\;
    \FOR{$b = \{1, \dots, B \}$}
        \STATE Sample a subset $U_b, y_b, f_b$ with $n$ points from $U, y, f$, respectively\; 
        \STATE Calculate ELBO of $\ell_\text{ELBO}$ based on \eqref{eq:obj-elbo} given data $U_b, y_b, f_b$\;
        \STATE Calculate the gradient of \eqref{eq:objective} {\it w.r.t.} $\theta$\;
        \STATE Calculate the natural gradient of \eqref{eq:objective} {\it w.r.t.} $V, m, S$\;
        \STATE Ascend the gradient of $\theta, V, m, S$\;
    \ENDFOR
\end{algorithmic}
\caption{Learning algorithm for the proposed model}
\label{algo:learning}
\end{algorithm}

\section{Derivation of predictive posterior}
\label{append:pred-posterior}

A Bayesian model makes predictions based on the posterior distribution. Given testing locations $U^*$, we can derive the predictive posterior distribution $p(\epsilon^*|y)$:
\begingroup
\allowdisplaybreaks
\begin{align*}
    p(\epsilon^*|y) 
    = &~ \int \int p(\epsilon^*, \epsilon, z|y) d \epsilon d z\\
    = &~ \int \int p(\epsilon^* |\epsilon, z, y) p(\epsilon, z | y) d \epsilon d z\\
    = &~ \int \int p(\epsilon^* |\epsilon, z) p(\epsilon, z | y) d \epsilon d z\\
    = &~ \int \int p(\epsilon^* |\epsilon, z) q(\epsilon, z) d \epsilon d z\\
    = &~ \int \int p(\epsilon^* |\epsilon, z) p(\epsilon | z) q(z) d \epsilon d z\\
    = &~ \int \left ( \int p(\epsilon^* |\epsilon, z) p(\epsilon | z) d \epsilon \right ) q(z) d z\\
    = &~ \int \left ( \int p(\epsilon^*, \epsilon | z) d \epsilon \right ) q(z) d z\\
    = &~ \int p(\epsilon^* | z) q(z) d z
\end{align*}
\endgroup

Similar to \eqref{eq:joint-dist-f-u}, we assume that the unobserved future data comes from the same generation process, i.e., 
\[
    p(\epsilon^*, z) = \mathcal{N}\left(~
    \begin{bmatrix}
    \epsilon^* \\
    z
    \end{bmatrix}
    ~\bigg|~
    \mathbf{0},~
    \begin{bmatrix}
    K_{**} & K_{*Z} \\
    K_{*Z}^\top & K_{*Z}
    \end{bmatrix}~
    \right),
\]
we can apply the multivariate Gaussian conditional rule on the prior $p(\epsilon^*, z)$ and obtain:
\vspace{-0.15in}
\[
    p(\epsilon^* | z)
    = ~\mathcal{N}(\epsilon^* | A_* z, B_*).
    \vspace{-0.15in}
\]
Combined with $q(z) = \mathcal{N}(m, S)$ we have
\vspace{-0.15in}
\[
    p(\epsilon^*|y) = ~\mathcal{N}(\epsilon^* | A_* m, A_* S A_*^\top + B_*)
    \vspace{-0.15in}
\]
where $A_* = K_{*Z}K_{VV}^{-1}$ and $B_* = K_{**} - K_{*Z} K_{VV}^{-1} K_{*Z}^\top$.



\vspace{0.1in}
\section{Experimental details}
\label{append:add-exp}

A detailed description of the baselines is provided as following:

\begin{enumerate}
    
    \item \emph{Neural network-based univariate prediction model} (\texttt{NN}) developed by \cite{CATALAO20111245} adopts a fully-connected neural network for wind speed prediction. The model is built on a three-layer neural network with a hidden layer size of six and a Hyperbolic activation function. To apply the method to our data, the model takes the first 12 hours of wind speed data as its input and generates the prediction for the next three hours. 

    \item \emph{Vector autoregressive} (\texttt{VAR}) model \citep{lutkepohl2013vector} is a generalization of the autoregressive (AR) model in multivariate scenarios. 
    Let the true wind average wind speed under data resolution $\xi_i$ at time $t$ be $y_i(t)$. In \texttt{VAR}($p$) model it is represented as 
    \begin{equation}
        \begin{aligned}
        y_i(t)= c + \sum_{\tau=1}^p A_\tau y_i(t-\tau)+\epsilon_{t},
        \end{aligned}
    \end{equation}
    where $c$ is the intercept vector of size $\xi_i^{\text{loc}}$, $p$ is the time lag, $A_{\tau}$ is coefficient matrix of size $\xi_i^{\text{loc}} \times \xi_i^{\text{loc}}$, and $\epsilon_{t}$ is the $\xi_i^{\text{loc}}$-dimensional error term. 
    In our experimental setting, we set $p=10$ for \texttt{VAR} model.

    \item \emph{Long Short-Term Memory} (\texttt{LSTM}) \citep{hochreiter1997long} is a recurrent neural network architecture that has been widely used to model time series data. 
    In this paper, we apply a sliding window of size $d$ on the raw data, which converts the input of LSTM to matrices of size $\xi_i^{\text{loc}} \times d$ and the output of LSTM to vectors of size $\xi_i^{\text{loc}} \times 1$. The input vector is passed into an LSTM unit of hidden size eight. 
    The output of LSTM is later passed into a dense layer to reduce the output dimension to 1.

    \item \emph{Global deep learning multivariate probabilistic forecasting models}, namely, \texttt{DeepAR} model \citep{SALINAS20201181} and \texttt{DeepTCN} model \citep{CHEN2020491} are developed for wind prediction. According to original papers, the input of these models are the past multivariate time series of wind variables with a size of $\xi_i^{\text{loc}} \times d$, and the future observations are assumed to come from a Gaussian distribution (referred to as DeepTCN-Gaussian in \cite{CHEN2020491}). We estimate the mean and variance parameters via the maximum likelihood estimator (MLE) for both \texttt{DeepAR} and \texttt{DeepTCN} models. The estimated mean of the Gaussian distributions is regarded as the deterministic forecast for wind speed.

\end{enumerate}

\end{document}